\documentclass{aa}
\usepackage{natbib,twoopt}
\usepackage{amsmath}

\usepackage{color}

\usepackage[breaklinks=true]{hyperref}
\makeatletter
  \newcommandtwoopt{\intheteads}[3][][]{\href{http://adsabs.harvard.edu/abs/#3}
    {\def\hyper@linkstart##1##2{}
     \let\hyper@linkend\@empty\citealp[#1][#2]{#3}}}
  \newcommandtwoopt{\citepads}[3][][]{\href{http://adsabs.harvard.edu/abs/#3}
    {\def\hyper@linkstart##1##2{}
     \let\hyper@linkend\@empty\citep[#1][#2]{#3}}}
  \newcommandtwoopt{\citetads}[3][][]{\href{http://adsabs.harvard.edu/abs/#3}
    {\def\hyper@linkstart##1##2{}
     \let\hyper@linkend\@empty\citet[#1][#2]{#3}}}
  \newcommandtwoopt{\citeyearads}[3][][]
    {\href{http://adsabs.harvard.edu/abs/#3}
    {\def\hyper@linkstart##1##2{}
     \let\hyper@linkend\@empty\citeyear[#1][#2]{#3}}}
\makeatother

\defcitealias{2022Natur.601..334Z}{Z+22}
\defcitealias{2024ApJ...973..136O}{O+24}
\defcitealias{2024A&A...685A..82E}{E+24}
\defcitealias{2024Natur.631...49S}{S+24}
\defcitealias{2025A&A...702A..12R}{Paper I}

\usepackage{graphicx}
\usepackage{txfonts}
\begin{document}

   \title{SISSI: Supernovae in a stratified, shearing interstellar medium}

   \subtitle{II. Star formation near the Sun is quenched by expansion of the Local Bubble}

   \author{Leonard E. C. Romano
          \inst{1}\fnmsep\inst{2}\fnmsep\inst{3}\fnmsep\thanks{Corresponding author: Leonard E. C. Romano\\\email{lromano@usm.lmu.de}}
          \and
          Andreas Burkert\inst{1}\fnmsep\inst{2}\fnmsep\inst{3}
          }

   \institute{Universitäts-Sternwarte, Fakultät für Physik, Ludwig-             Maximilians-Universität München, Scheinerstr. 1, D-               81679 München, Germany
         \and
             Max-Planck-Institut für extraterrestrische Physik, Giessenbachstr. 1, D-85741 Garching, Germany
        \and
             Excellence Cluster ORIGINS, Boltzmannstr. 2, D-85748 Garching, Germany
             }
 
  \abstract 
   {The age of the Local Bubble (LB) constrains the timescale on which the interstellar medium in the solar neighborhood evolves. 
   Previous estimates placed the age of the LB at $\gtrsim 14\,\text{Myr}$, and attributed its expansion to $\sim 15-20$ supernovae (SNe), yet a companion paper suggests this age may be overestimated.
   }
   {We place new constraints on the age of the LB and re-evaluate the question whether its expansion triggered or suppressed local star formation.}
   {We reconstruct the LB’s geometry and momentum using publicly available 3D dust maps and compare them to the high-quality sample of simulated supernova remnants in the SISSI project. Independent constraints on the star-formation history and supernova rate are obtained from a Gaia DR3–based census of nearby star clusters.
   }
   {
   We find that $\sim7- 59$ SNe over $\sim5.8\,\text{Myr}$ to $\sim2.8\,\text{Myr}$, respectively,
   are required to explain both the LB's momentum and size and confirm that such a high supernova rate can be sustained by local star clusters.
   } 
   {
   Our analysis yields a substantially smaller LB age than previous estimates, requiring a correspondingly larger number of SNe, driving its expansion. We show that this result is in tension with the conclusion that the LB is powered solely by SNe from the Scorpius–Centaurus OB association, which ceased star formation around the time the LB formed.
   If our estimates are correct, it follows that the majority of star formation in the solar neighborhood happened before the formation of the LB and was not triggered by its expansion. Instead, the SNe that powered the LB appear to overall have quenched the ongoing star formation process. This does not rule out that star formation in the clouds, located near its current edge, could have been affected by the LB expansion.
   }


   \maketitle

\section{Introduction}
Recent advances in observational methods allow us to map out the structure in the nearby galactic interstellar medium (ISM) in three spatial dimensions (3D) \citep[3D, e.g.,][]{1992A&A...258..104A, 2019A&A...625A.135L, 2024A&A...685A..82E}. 
One of the most prominent structures, mapped by these techniques is the Local Bubble \citep[hereafter LB][]{1987ARA&A..25..303C, 2021ApJ...920...75L}, a large cavity, several hundred parsec across, centered around the solar system \citep[][hereafter Z+22]{2022Natur.601..334Z}, which has been linked to diffuse soft X-ray emission \citep[e.g.,][]{2000ApJS..128..171S, 2024A&A...690A.399Y}.
Superbubbles (SB) such as the LB are believed to be carved out by the various feedback channels of massive stars, such as stellar winds \citep{1990MNRAS.244..563T}, ionizing radiation \citep{2021ApJ...920...75L} and supernovae \citep[hereafter SNe][]{1990MNRAS.244..563T, 2006A&A...452L...1B, 2021Sci...372..742W}.

Over the past few decades, many studies have tried to constrain the origin of the LB. 
Studies using astrometry, tracing back the positions of nearby star-clusters suggest that $\sim 10-20$ SNe originating from the Scorpius-Centaurus OB association (Sco-Cen) might have contributed to the expansion of the LB $\sim 10\,\text{Myr}$ ago \citep{2001ApJ...560L..83M, 2006MNRAS.373..993F, 2022Natur.601..334Z}. 
Numerical simulations of SNe expanding into a turbulent, stratified ISM, confirm that $\sim 20$ SNe exploding sequentially throughout the last $\sim 14\,\text{Myr}$ could explain, the observed size of the LB, the observed column densities of \ion{O}{VI} in the interior as well as the deposition of sedimentary radioisotopes, such as $^{60}$Fe \citep{2006A&A...452L...1B, 2016Natur.532...73B, 2023A&A...680A..39S}. 
Fossil records show that the incorporation rate of $^{60}$Fe and $^{244}$Pu on earth has peaked $\lesssim 4\, \text{Myr}$ and $\lesssim 7\,\text{Myr}$ ago \citep{2016Natur.532...69W, 2021Sci...372..742W}.
Combining this with the work of \citetalias{2022Natur.601..334Z}, who find that the solar system would have entered into the LB $\sim 5\,\text{Myr}$ ago, suggests that at least the more recent peak might coincide with SNe associated with the LB \citep[see also][]{2016Natur.532...73B}.

In \citep[][hereafter Paper I]{2025A&A...702A..12R}, we introduce the SISSI (Supernovae In a Stratified, Shearing ISM) zoom-in simulation project, focusing on the properties of highly-resolved, simulated supernova-remnants (SNRs) expanding into the self-consistently generated ISM of an isolated, Milky-Way-like disk-galaxy.
In \citetalias{2025A&A...702A..12R} we find that SNRs in a realistic environment, expand faster than previously expected at later times $t\gtrsim 1\,\text{Myr}$, due to galactic shear, the gravitational influence of nearby substructures and the presence of low-density channels. 
A comparison of our simulated sample of SNRs at $t = 10\,\text{Myr}$ with the LB, suggested that it would be too small for a SB of its age, powered by SNe exploding at the rate suggested by previous studies.

In this work, we follow up on this curiosity, by presenting a new analysis of the LB, derived from the 3D dust maps of \citet{2024A&A...685A..82E} (hereafter E+24) using a method similar to that of \citet{2024ApJ...973..136O} (hereafter O+24).
As opposed to \citetalias{2022Natur.601..334Z}, who assume that the stellar kinematics trace the dynamics of the gas through triggered star-formation, we directly estimate the size, momentum and ambient density of the LB from the 3D dust maps of \citetalias{2024A&A...685A..82E} and by utilizing the dynamical evolution of the SNRs in the SISSI simulation (see Appendix \ref{app:SISSI} for a brief description of the simulation suite)
we constrain its age and the number of SNe powering its evolution.

\section{The age of the LB} \label{sec:SISSI}
\begin{figure}
 \centering
 \includegraphics[width=0.8\linewidth, clip=true]{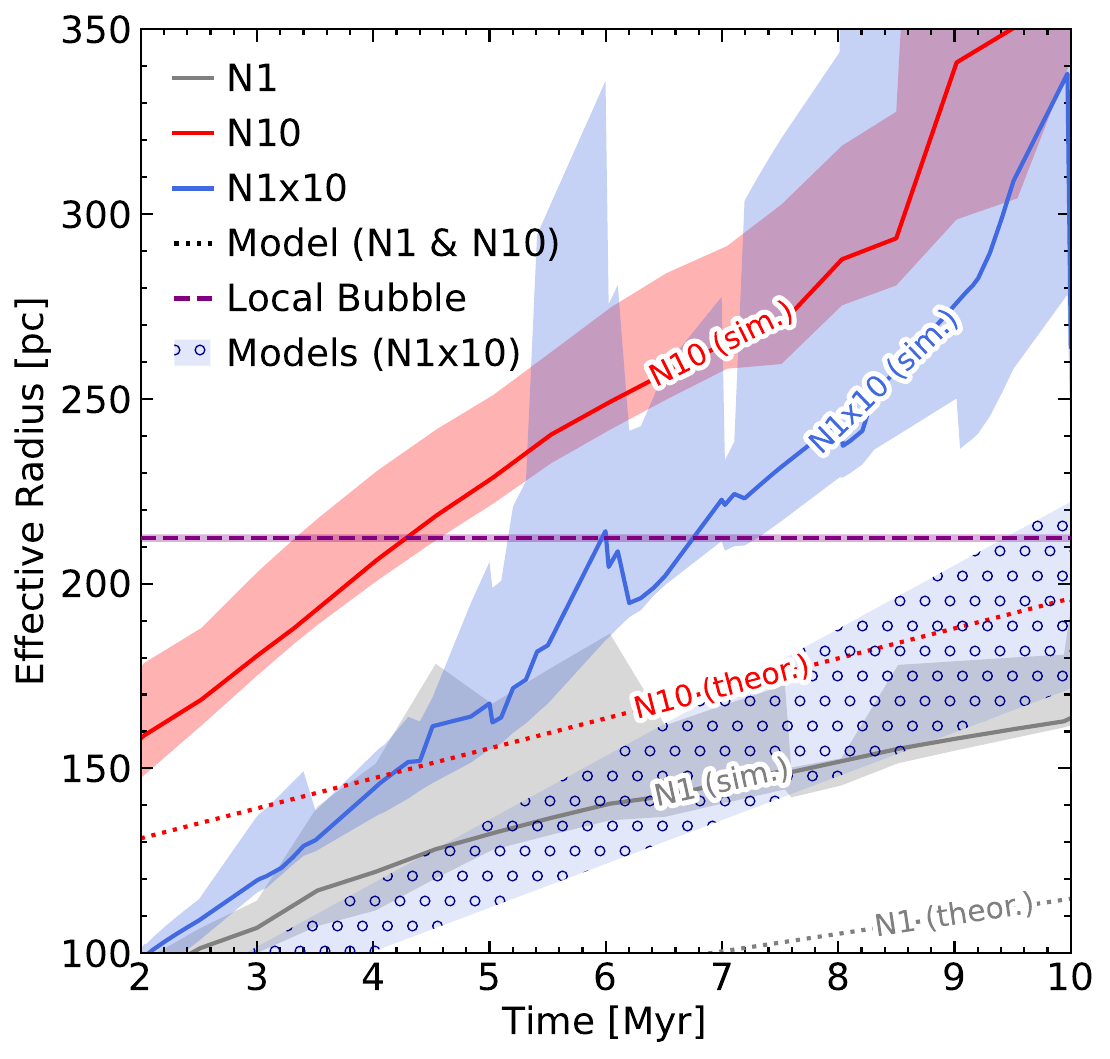}
 \caption{Time evolution of effective size of the simulated sample of SNRs of the SISSI sample between 2 and 10 Myr, for SNRs with ambient densities within 0.3 dex of the LB (Tab. \ref{tab:local_bubble}).
 Gray, red and blue lines correspond to different explosion models. Solid lines correspond to the median size of the simulated bubbles, with shaded areas corresponding to the range between the 30th and 70th percentiles. 
 Dotted lines and the hatched blue contour correspond to theoretical models based on 
 radiative blastwaves in uniform media (Appendix~\ref{app:SB_models}).
 For model N10, the size of the LB corresponds to an age of $\sim 4.5\,\text{Myr}$, while for N1x10 it corresponds to $\sim 6-7\,\text{Myr}$.} 
 \label{fig:size_evolution}
\end{figure}

The properties of a SB are determined by the strength of its energy source, the density of the environment and its age.
Previous studies suggest that the radial momentum imparted per SN onto an expanding radiative shell only depends weakly on the density and is largely independent of the age of the SB \citep{2015MNRAS.451.2757W, 2022ApJS..262....9O}, namely, by knowing the momentum, we can precisely determine the number of SNe responsible for the expansion of the SB.
In \citetalias{2025A&A...702A..12R} we have found that the average momentum imparted per SN is approximately
\begin{equation} \label{eq:momentum_per_SN}
    \hat{p}_{\text{SN}} \sim  \left(2.6 \pm 0.4\right) \times 10^5 \, n_{0}^{-0.13} \, \text{M}_{\odot}\,\text{km s}^{-1} ~,
\end{equation}
where $n_{\text{H}} = n_{0}\,\text{cm}^{-3}$ is the ambient number density of hydrogen and we estimate the systematic uncertainty based on the range of values found in numerical studies \citep[e.g.][]{2015MNRAS.451.2757W, 2015ApJ...802...99K, 2015MNRAS.450..504M}, dependent on the detailed cooling physics, resolution and other numerical choices. In a purely energy-driven model that can arise in situations where cooling is inefficient \citep{1998LNP...506..483K}, $\hat{p}_{\text{SN}}$ increases slightly with time \citep{2019MNRAS.490.1961E, 2022Natur.601..334Z}, though for the range of ages considered in the SISSI simulations these differences are minor.

In App. \ref{app:local_bubble} we find the LB's average ambient density of $\sim 0.5 \, \text{cm}^{-3}$, an age t$_{age}$ dependent momentum of $\sim 3 \times 10^{7} \,(\text{Myr}/t_{\text{age}} \, \text{M}_{\odot} \, \text{km s}^{-1}$), a mass of $\sim 6.2\times10^{5}\, \text{M}_{\odot}$ and a size of $\sim 212.3\,\text{pc}$ (see Tab. \ref{tab:local_bubble} for uncertainties and a comparison with previous estimates).
Using these estimates we derive a momentum input per SN of $\hat{p}_{\text{SN, LB}} \sim \left(2.9 \pm 0.5_{\text{sys}} \pm 0.005_{\text{stat}}\right) \times 10^5\, \text{M}_{\odot}\,\text{km s}^{-1}$.
We related the number of SNe to its age by utilizing the age-dependent momentum estimate
\begin{align}\label{eq:N_SN}
    N_{\text{SN, LB}}\left(t_{\text{age}}\right) &= p_{\text{LB}} / \hat{p}_{\text{SN, LB}} \nonumber \\ &\sim \left(104 \pm 2_{\text{stat}} \pm 59_{\text{sys}}\right)  \times \left(t_{\text{age}} / \text{Myr}\right)^{-1} ~.
\end{align}
The corresponding average SN rate is obtained by dividing by the age $\dot{N}_{\text{SN, LB}} = N_{\text{SN, LB}} / t_{\text{age}}$.
We validated the applicability of this estimate in Appendix~\ref{app:N_SN}.

In contrast to the momentum, the size of a SB depends only weakly on the energy input, and in turn depends most strongly on its age. 

In Fig. \ref{fig:size_evolution} we show the time evolution of the effective size of our simulated sample of SNRs, introduced in \citetalias{2025A&A...702A..12R}, for SNRs with ambient densities within 0.3 dex of that of the LB. 
The SB evolving in a realistic galactic environment (solid lines) grow significantly faster than expected from blastwave models in uniform environments \citep[see Appendix~\ref{app:SB_models}][]{2019MNRAS.490.1961E, 2022ApJS..262....9O}.
The size of the SB driven by subsequent SNe (N1x10) closely matches that of the SNR of a single SN (N1) for $t \lesssim 3\,\text{Myr}$, while it approaches that of the SNR of 10 SNe exploding all at once (N10) after $\sim 10\,\text{Myr}$.
For the models with multiple SNe, the size of the LB is reached after $\sim 4.5\,\text{Myr}$ (N10) and after $6-7\,\text{Myr}$ (N1x10). 

At the age at which the simulated SNRs (N1x10) reach the size of the LB, $17 \pm 10_{\text{sys}}$ SNe are required to explain its observed momentum (Eq. \ref{eq:N_SN}), in contrast to the 6-7 SNe that exploded in the simulation. The difference arises from the fact that the simulations were not fine tuned in order to match the observationally based momentum of the LB. A SB powered by $\sim 17$ SNe would have reached the same size even earlier than in the simulation making the difference even large in LB age between our evaluation and previous age estimates \citep{2001ApJ...560L..83M, 2006A&A...452L...1B, 2022Natur.601..334Z}. While the simulations are not fine-tuned to the observation one can get a rough estimate of the difference using theoretical considerations.

\begin{figure}
 \centering
 \includegraphics[width=0.8\linewidth, clip=true]{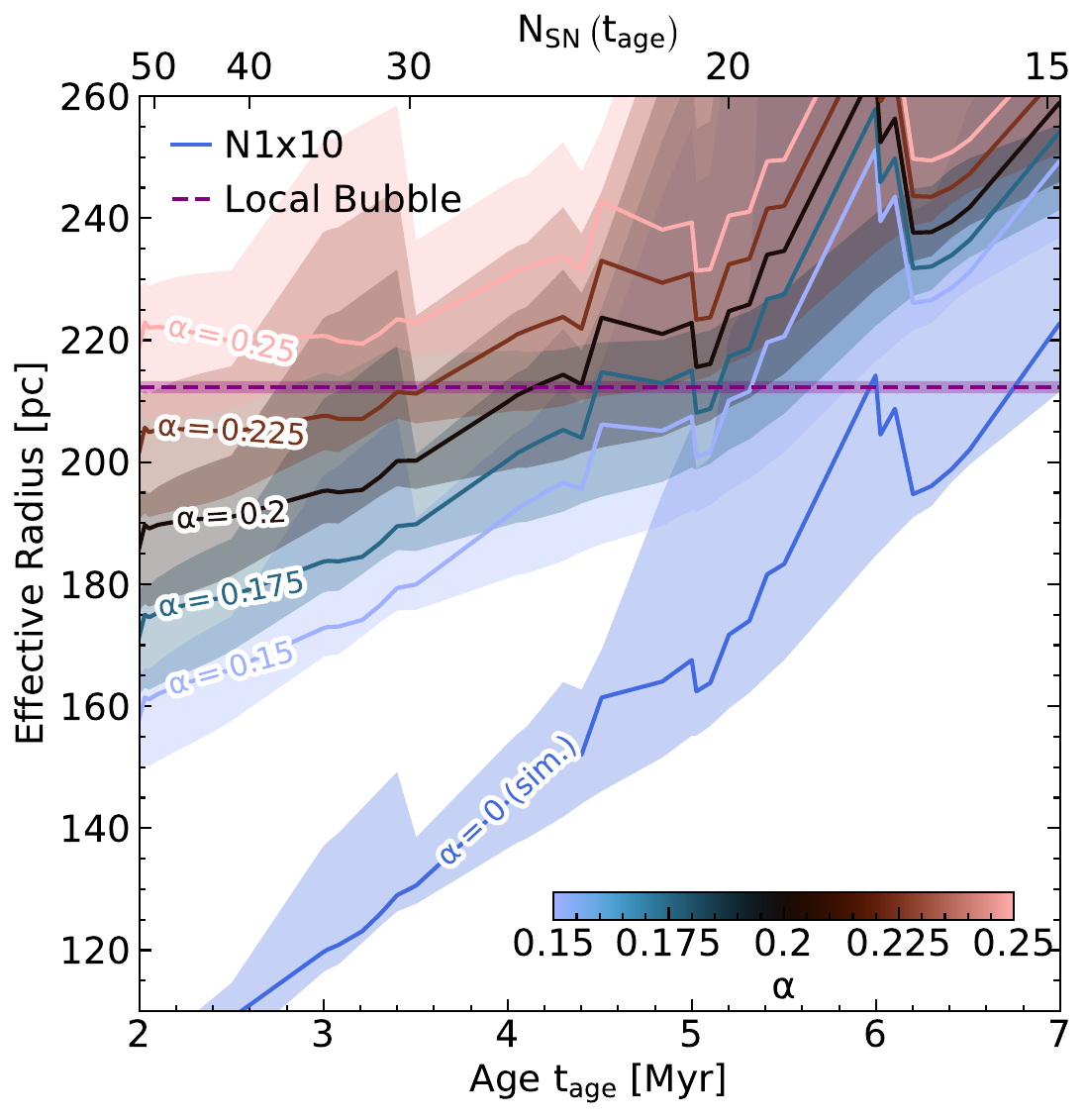}
 \caption{Effective size as a function of age / number of SNe extrapolated from the model N1x10 and the mean of the momentum constraint Eq. \ref{eq:N_SN} for various values of $\alpha$, namely, $R_{\alpha} = R_{\text{N1x10}}\left(t = t_{\text{age}}\right) \times \left[\dot{N}_{\text{SN, LB}}\left(t_{\text{age}}\right) \,  / \, 1 \, \text{Myr}^{-1}\right]^\alpha$, where $\dot{N}_{\text{SN, LB}} = N_{\text{SN, LB}} / t_{\text{age}}$.
 For $\alpha \lesssim 0.225$ the extrapolated sizes match the LB for ages in the range $3.5 - 5.5 \, \text{Myr}$ while for $\alpha \gtrsim 0.225$ the two constraints on the size and momentum cannot be satisfied simultaneously.} 
 \label{fig:size_extrapolation}
\end{figure}

Previous models of radiative blastwaves in uniform ambient media \citep{2019MNRAS.490.1961E, 2022ApJS..262....9O, 2024ApJ...970...18L} suggest that the effective size is roughly $\propto \dot{N}_{\text{SN}}^{0.15-0.25}$.
A blastwave powered by more frequent SNe, would thus only lead to a mild increase in the size, and would be largely consistent with an age $\lesssim 6\,\text{Myr}$.
In Fig. \ref{fig:size_extrapolation} we show that for $R \propto \left\langle\dot{N}_{\text{SN}}\right\rangle^{\alpha}$ with $\alpha \lesssim 0.225$ the size matches that of the LB for $t_{\text{age}} \sim 3.5-5.5\,\text{Myr}$, while for $\alpha >0.225$, there are no solutions that can simultaneously explain the size and the momentum of the LB. Not shown are scenarios with extreme $\dot{N}_{\text{SN}}$, but we confirm that they reproduce a similar range of ages ($\sim 2.8 - 5.8\, \text{Myr}$) and correspondingly lower and higher numbers of SNe ranging from $\sim7$ to $\sim59$ SNe.
A more detailed evaluation is out of the scope of this work and will be subject to future investigation (Romano et al., in prep.).

By invoking standard assumptions about the stellar initial mass function, the average time between SNe in a star cluster can be linked to its mass \citep{2017ApJ...834...25K}
\begin{equation}
    \Delta t_{\text{SN}} \sim 0.4 \, \left(\frac{M_{\text{cl}}}{10^4\,\text{M}_{\odot}}\right)^{-1} \, \text{Myr} ~,
\end{equation}
which for the range of supernova rates found above would correspond to a mass of $M_{\text{cl, LB}} \sim 0.5 - 8.4\,\times10^4\,\text{M}_{\odot}$ for the progenitor cluster of the LB.

We compared this number to the swept-up mass, and assuming that the swept up mass roughly corresponds to the mass of the birth-cloud of the cluster driving the expansion of the LB (i.e. neglecting the background ISM), obtained an upper limit for the cloud-scale star-formation efficiency
\begin{equation}
    \epsilon_{\star} = \frac{M_{\text{cl, LB}}}{M_{\text{sw, LB}}} \lesssim 0.8-14\,\%~, 
\end{equation}
in agreement with the range of values found in observational \citep{2016ApJ...833..229L, 2020MNRAS.493.2872C} and numerical studies \citep{2022MNRAS.512..216G, 2024MNRAS.527.6732F}.

We summarize our results and compare them to those of \citetalias{2022Natur.601..334Z} in Tab. \ref{tab:SNe}.
The obtained cluster mass is somewhat higher than the stellar mass of Sco-Cen \citep{2018A&A...619A.120K}, which is oftentimes assumed to be the progenitor cluster of the LB \citep{2001ApJ...560L..83M, 2022Natur.601..334Z}.
Similarly, our result that the LB should be younger than previously believed is in tension with the older age ($\sim 15-17 \, \text{Myr}$) of the stars in Sco-Cen, indicating that Sco-Cen stars alone might not alone power the LB.

An additional constraint for the SN origin of the LB comes from the fossil records of sedimentary radionuclides \citep{2021Sci...372..742W, 2023ApJ...947...58E, 2024ApJ...972..179E}, which requires that the solar system entered the LB at least $\gtrsim 4\, \text{Myr}$ ago. Both the scenario proposed in this work (Appendix~\ref{app:solar_system}), and that of \citetalias{2022Natur.601..334Z} satisfy this constraint. A more detailed analysis of the dynamical incorporation of the dusty, radioactive material may provide additional evidence in favor of either scenario. 

\begin{table}[t]
\caption{\centering Expansion parameters of the LB.}
\label{tab:SNe}
\begin{center}
\resizebox{0.9\linewidth}{!}{
\begin{tabular}{l|l l l|l}
  Property & \multicolumn{2}{c}{This work}  & \citetalias{2022Natur.601..334Z} & Unit\\
   & High $N_{\text{SN}}$ & Low $N_{\text{SN}}$ & & \\\hline
   Number of SNe & $\lesssim 59$ & $\gtrsim 7 $ & $15^{+11}_{-7}$ & --\\
  Age & $\gtrsim2.8$ & $\lesssim 5.8 $ & $14.39^{+0.78}_{-0.74}$ & Myr\\
  Time between SNe & $\gtrsim0.05$ & $\lesssim0.8$ & $1.06^{+0.63}_{-0.39}$ & Myr\\
  Star cluster mass & $\lesssim 8.4$ & $\gtrsim0.5$ & $\sim 0.4$ & $10^4 \, \text{M}_{\odot}$\\
  Star-Formation Efficiency & $\lesssim 14$ & $\gtrsim0.8 $ & $\sim 0.3$ & $\%$ \\
  Expansion speed\tablefootmark{a} & $\lesssim 37^{+21}_{-10} $ & $ \gtrsim18^{+10}_{-5}$ & $6.7^{+0.5}_{-0.4}$& km s$^{-1}$\\
  \hline      
\end{tabular}
}
\end{center}
\begin{center}
\resizebox{0.9\linewidth}{!}{
\tablefoottext{a}{Central value computed from $R_{\text{eff}}$, lower (upper) limits from 
minor (major) axis} 
}
\end{center}
\end{table}

According to \citet{2024Natur.631...49S} (hereafter, S+24), Sco-Cen is part of a larger family of stars known as the $\alpha$-Persei family ($\alpha$Per), which reportedly had an average time between SNe of $\Delta t_{\alpha\text{Per}} = 0.32^{+0.50}_{-0.23} \,\text{Myr}$.
Moreover, the star-formation history (SFH) of $\alpha$Per indicates episodic star-formation over the past $\gtrsim60\,\text{Myr}$ with peaks every $\gtrsim 12.5\,\text{Myr}$ and the latest peak within the last $\sim 12.5\,\text{Myr}$.

Our results are in agreement with the SFH in $\alpha$Per if we assume that the LB was powered by the latest peak of star-formation rather than the trough associated with the formation of Sco-Cen.
In Appendix \ref{app:AlphaPersei} we constructed a more detailed star-formation and SN-rate history over the past 40 Myr to substantiate this claim.
The SN rate and SFH resulting from these considerations are shown in Fig. \ref{fig:SFH}. The upper panel shows that the peak of star formation in the LB area is before the onset of SNe that fuel its expansion, followed by a sudden distinct downturn, coincident with the range of ages found here. This suggests that the expansion of the LB may have predominantly quenched star formation in the solar neighborhood, rather than triggering it. While LB-triggered star-formation is not entirely ruled out, especially when one considers young star-forming regions the LB edge, it appears to be subdominant.

In Fig. \ref{fig:trajectories} we show the orbits of the clusters potentially contributing SNe to the LB. We find that while the majority of the active clusters are coincident with Sco-Cen, in line with the expectation that the LB likely originated from Sco-Cen \citep{2006MNRAS.373..993F}, there were likely significant contributions from other stellar associations in the solar neighborhood.

The high expansion speed $v_{\text{LB}} \sim 0.5 \, R_{\text{LB}} / t_{\text{age}} \gtrsim 10\,\text{km s}^{-1}$ (Tab. \ref{tab:SNe}) suggests that the LB might be just above the regime, where it is starting to be affected by its Galactic environment, which typically dominates as $v \lesssim \sigma_{\text{turb}} \sim 10\,\text{km s}^{-1}$ \citepalias{2025A&A...702A..12R}.
Thus, models for SBs expanding into a uniform medium (Appendix~\ref{app:SB_models}) might provide a reasonable age-estimate after all.
For the slowly-cooling-wind model used by \citetalias{2022Natur.601..334Z} with updated ambient density and the SN rate following \citetalias{2024Natur.631...49S} we obtained $t_{\text{age}}^{\text{SCW}} \sim 6.6^{+2.4}_{-2.3}\,\text{Myr}$, while for a rapidly-cooling-wind model \citep{2022ApJS..262....9O} we obtained $t_{\text{age}}^{\text{RCW}} \sim 9^{+5}_{-4}\,\text{Myr}$, in rough agreement with both scenarios.

\citet{2012A&A...539L...1D} compare the local ISM simulations of the LB and the adjacent Loop I SB to the observed ion abundances of \ion{C}{iv}, \ion{N}{v} and \ion{O}{vi} to constrain the expansion history of the LB. They find, that within the scope of their model of the local ISM, the last SN explosion within the LB must have occurred $0.5 - 0.8\,\text{Myr}$ ago. These numbers are in rough agreement with the lower end of SN rates found here (low N$_{SN}$ column of Tab. \ref{tab:SNe}). They are somewhat larger than the SN rates, based on recent Gaia data \citepalias[$\sim 0.32^{+0.50}_{-0.23} \,\text{Myr}$,][]{2024Natur.631...49S}.
One reason for this discrepancy could be the slightly higher average ambient density of $\sim 1\,\text{cm}^{-3}$ in the simulations  
of \citet{2012A&A...539L...1D}, leading to shorter overall chemistry timescales. In a lower density environment the same constraint might be thus in better agreement with the higher SN rates found here. A detailed comparison is out of the scope of this work and will be subject to future investigation.

The LB appears to be at the tipping point between microscopic SNRs, whose dynamics are dominated by local shock physics and mesoscopic SBs that are strongly affected by galactic-scale processes.
More detailed models of such large SBs as well as more precise stellar ages \citep[e.g.,][]{2022A&A...667A.163M, 2023A&A...678A..71R, 2024Natur.631...49S} are needed to shed light on the progenitor cluster and expansion history of the LB.

\section{Concluding remarks}\label{sec:conclusion}

We compared the momentum and the effective size of the LB, derived from publicly available, high-quality 3D dust maps to our sample of simulated SNRs expanding into the shearing, stratified ISM of the isolated Milky-Way-like SISSI galaxy.

We find that in order to match both constraints on the radial momentum, and the effective size, the LB would have to be driven by a larger number of SNe and be significantly younger than previously reported.
In particular, a rough agreement requires an age since the onset of SNe of $ t_{\text{age, LB}} \sim 2.8-5.8\,\text{Myr} $ for $N_{\text{SN, LB}} \sim 7 - 59$,
in agreement with estimates of the average time between SNe in $\alpha$Per \citepalias{2024Natur.631...49S}. 
Investigating the star formation history of the solar neighborhood (Fig. \ref{fig:SFH}), such a young age indicates that SNe feedback induced formation of the LB overall quenched local star-formation although new clouds forming at its shell might now trigger new generations \citepalias{2022Natur.601..334Z}. Another possibility is that the nearby star-forming regions (e.g., Sco-Cen, Taurus) were already formed before the formation of the LB. Indeed, these regions are all closer \citep{2021ApJ...919...35Z} than the effective radius of the LB ($\sim 210 \, \text{pc}$), indicating that they might be entrained by it \citep[see e.g., Fig. 13 of ][]{2024ApJ...965..168R}.  This might trigger further star formation. While our current methodology, which defines the bubble's boundary as the first prominent density peak along each line of sight, cannot disentangle these detailes , a more sophisticated method (e.g., O’Neill et al., EAS 2025) might provide more insight.

We stress that the conclusions of this work that star formation near the sun might be quenched and driven \citepalias{2022Natur.601..334Z} by the expansion of the LB are not mutually exclusive. Star-formation on large scales can be quenched while locally the conditions for new star-formation are triggered. The degree to which SN-driven SBs, such as the LB, might trigger and quench star-formation on large ($\gtrsim 100 \, \text{pc}$) scales is a highly interesting question that is not well understood.
The SISSI simulations are well suited for this task and an investigation of this pertinent question is currently underway (Romano et al., in prep.).

We also not that while our analysis focuses on SNe only, a fraction of the energy is also provided from other sources such as stellar winds and \ion{H}{II} regions.
Though, it is worth stressing that \citet{2023A&A...680A..39S} consider the role of stellar winds and find that before the onset of SNe, the LB maintains an approximately constant size of $\sim 30\, \text{pc}$ for $\sim 10 \, \text{Myr}$ suggesting that the contribution from winds might be minor. In this context the age of the LB could be inflated by prolonged periods of negligible growth. We do not count such prolonged periods of halted growth towards the age of the LB.

While our results are in tension with previous estimates, which assumed that the LB was exclusively powered by SNe in Sco-Cen, they are in better agreement with more recent observational data on the more global star formation in the solar neighborhood. We have used scaling relations in order to extrapolate our simulations to the unique conditions of the LB. More detailed models for old ($t \gtrsim 1\,\text{Myr}$) SBs expanding into a realistic galactic ISM as well as additional observational constraints are now required to resolve remaining uncertainties and obtain a clearer picture for the origin and history of the LB.

\begin{acknowledgements}
We thank Manuel Behrendt for detailed and very fruitful discussions and comments that substantially improved the quality of the manuscript and for providing an early version of the AVALON galaxy simulation code which is the basis for the SISSI simulations. We also thank the anonymous referee for their insightful comments and suggestions that helped to improve the quality of this work.
This research was funded by the Deutsche Forschungsgemeinschaft (DFG, German Research Foundation) under Germany's Excellence Strategy – EXC 2094 – 390783311.
LR thanks the developers of the following software and
packages that were used in this work: \textsc{Julia} v1.10.0 \citep{Julia-2017},
\textsc{Matplotlib} v3.5.1 \citep{Hunter:2007},
\textsc{Healpix} v2.3.0 \citep{2021ascl.soft09028T}, and \textsc{Galpy} v.1.10.2 \citep{2015ApJS..216...29B}
The Julia source-code for our analysis is made available at \href{https://doi.org/10.5281/zenodo.17054923}{https://doi.org/10.5281/zenodo.17054923}
\end{acknowledgements}

\bibliographystyle{aa}
\bibliography{bibliography}

@ARTICLE{2024ApJ...973..136O,
       author = {{O'Neill}, Theo J. and {Zucker}, Catherine and {Goodman}, Alyssa A. and {Edenhofer}, Gordian},
        title = "{The Local Bubble Is a Local Chimney: A New Model from 3D Dust Mapping}",
      journal = {\apj},
         year = 2024,
        month = oct,
       volume = {973},
       number = {2},
          eid = {136},
        pages = {136},
          doi = {10.3847/1538-4357/ad61de},
archivePrefix = {arXiv},
       eprint = {2403.04961},
 primaryClass = {astro-ph.GA},
       adsurl = {https://ui.adsabs.harvard.edu/abs/2024ApJ...973..136O}
}

@ARTICLE{2024A&A...685A..82E,
       author = {{Edenhofer}, Gordian and {Zucker}, Catherine and {Frank}, Philipp and {Saydjari}, Andrew K. and {Speagle}, Joshua S. and {Finkbeiner}, Douglas and {En{\ss}lin}, Torsten A.},
        title = "{A parsec-scale Galactic 3D dust map out to 1.25 kpc from the Sun}",
      journal = {\aap},
         year = 2024,
        month = may,
       volume = {685},
          eid = {A82},
        pages = {A82},
          doi = {10.1051/0004-6361/202347628},
archivePrefix = {arXiv},
       eprint = {2308.01295},
 primaryClass = {astro-ph.GA},
       adsurl = {https://ui.adsabs.harvard.edu/abs/2024A&A...685A..82E}
}

@article{Julia-2017,
    title={Julia: A fresh approach to numerical computing},
    author={Bezanson, Jeff and Edelman, Alan and Karpinski, Stefan and Shah, Viral B},
    journal={SIAM {R}eview},
    volume={59},
    number={1},
    pages={65--98},
    year={2017},
    publisher={SIAM},
    doi={10.1137/141000671},
    url={https://epubs.siam.org/doi/10.1137/141000671}
}

@Article{Hunter:2007,
  Author    = {Hunter, J. D.},
  Title     = {Matplotlib: A 2D graphics environment},
  Journal   = {Computing in Science \& Engineering},
  Volume    = {9},
  Number    = {3},
  Pages     = {90--95},
  publisher = {IEEE COMPUTER SOC},
  doi       = {10.1109/MCSE.2007.55},
  year      = 2007
}

@MISC{2021ascl.soft09028T,
       author = {{Tomasi}, Maurizio and {Li}, Zack},
        title = "{Healpix.jl: Julia-only port of the HEALPix library}",
      version = {2.3.0},     
         year = 2021,
        month = sep,
          eid = {ascl:2109.028},
        pages = {ascl:2109.028},
archivePrefix = {ascl},
       eprint = {2109.028},
       adsurl = {https://ui.adsabs.harvard.edu/abs/2021ascl.soft09028T}
}

@ARTICLE{2015MNRAS.451.2757W,
       author = {{Walch}, Stefanie and {Naab}, Thorsten},
        title = "{The energy and momentum input of supernova explosions in structured and ionized molecular clouds}",
      journal = {\mnras},
         year = 2015,
        month = aug,
       volume = {451},
       number = {3},
        pages = {2757-2771},
          doi = {10.1093/mnras/stv1155},
archivePrefix = {arXiv},
       eprint = {1410.0011},
 primaryClass = {astro-ph.GA},
       adsurl = {https://ui.adsabs.harvard.edu/abs/2015MNRAS.451.2757W}
}

@ARTICLE{2022ApJS..262....9O,
       author = {{Oku}, Yuri and {Tomida}, Kengo and {Nagamine}, Kentaro and {Shimizu}, Ikkoh and {Cen}, Renyue},
        title = "{Osaka Feedback Model. II. Modeling Supernova Feedback Based on High-resolution Simulations}",
      journal = {\apjs},
         year = 2022,
        month = sep,
       volume = {262},
       number = {1},
          eid = {9},
        pages = {9},
          doi = {10.3847/1538-4365/ac77ff},
archivePrefix = {arXiv},
       eprint = {2201.00970},
 primaryClass = {astro-ph.GA},
       adsurl = {https://ui.adsabs.harvard.edu/abs/2022ApJS..262....9O}
}

@ARTICLE{2024ApJ...965..168R,
       author = {{Romano}, Leonard E.~C. and {Behrendt}, Manuel and {Burkert}, Andreas},
        title = "{Cloud Formation by Supernova Implosion}",
      journal = {\apj},
         year = 2024,
        month = apr,
       volume = {965},
       number = {2},
          eid = {168},
        pages = {168},
          doi = {10.3847/1538-4357/ad2c05},
archivePrefix = {arXiv},
       eprint = {2402.05796},
 primaryClass = {astro-ph.GA},
       adsurl = {https://ui.adsabs.harvard.edu/abs/2024ApJ...965..168R}
}

@ARTICLE{2022Natur.601..334Z,
       author = {{Zucker}, Catherine and {Goodman}, Alyssa A. and {Alves}, Jo{\~a}o and {Bialy}, Shmuel and {Foley}, Michael and {Speagle}, Joshua S. and {Gro{\^I}{\texttwosuperior}schedl}, Josefa and {Finkbeiner}, Douglas P. and {Burkert}, Andreas and {Khimey}, Diana and {Swiggum}, Cameren},
        title = "{Star formation near the Sun is driven by expansion of the Local Bubble}",
      journal = {\nat},
         year = 2022,
        month = jan,
       volume = {601},
       number = {7893},
        pages = {334-337},
          doi = {10.1038/s41586-021-04286-5},
archivePrefix = {arXiv},
       eprint = {2201.05124},
 primaryClass = {astro-ph.GA},
       adsurl = {https://ui.adsabs.harvard.edu/abs/2022Natur.601..334Z}
}

@ARTICLE{2021Sci...372..742W,
       author = {{Wallner}, A. and {Froehlich}, M.~B. and {Hotchkis}, M.~A.~C. and {Kinoshita}, N. and {Paul}, M. and {Martschini}, M. and {Pavetich}, S. and {Tims}, S.~G. and {Kivel}, N. and {Schumann}, D. and {Honda}, M. and {Matsuzaki}, H. and {Yamagata}, T.},
        title = "{$^{60}$Fe and $^{244}$Pu deposited on Earth constrain the r-process yields of recent nearby supernovae}",
      journal = {Science},
         year = 2021,
        month = may,
       volume = {372},
       number = {6543},
        pages = {742-745},
          doi = {10.1126/science.aax3972},
       adsurl = {https://ui.adsabs.harvard.edu/abs/2021Sci...372..742W}
}

@ARTICLE{2017ApJ...834...25K,
       author = {{Kim}, Chang-Goo and {Ostriker}, Eve C. and {Raileanu}, Roberta},
        title = "{Superbubbles in the Multiphase ISM and the Loading of Galactic Winds}",
      journal = {\apj},
         year = 2017,
        month = jan,
       volume = {834},
       number = {1},
          eid = {25},
        pages = {25},
          doi = {10.3847/1538-4357/834/1/25},
archivePrefix = {arXiv},
       eprint = {1610.03092},
 primaryClass = {astro-ph.GA},
       adsurl = {https://ui.adsabs.harvard.edu/abs/2017ApJ...834...25K}
}

@ARTICLE{2006A&A...452L...1B,
       author = {{Breitschwerdt}, D. and {de Avillez}, M.~A.},
        title = "{The history and future of the Local and Loop I bubbles}",
      journal = {\aap},
         year = 2006,
        month = jun,
       volume = {452},
       number = {1},
        pages = {L1-L5},
          doi = {10.1051/0004-6361:20064989},
archivePrefix = {arXiv},
       eprint = {astro-ph/0604162},
 primaryClass = {astro-ph},
       adsurl = {https://ui.adsabs.harvard.edu/abs/2006A&A...452L...1B}
}

@ARTICLE{2001ApJ...560L..83M,
       author = {{Ma{\'\i}z-Apell{\'a}niz}, Jes{\'u}s},
        title = "{The Origin of the Local Bubble}",
      journal = {\apjl},
         year = 2001,
        month = oct,
       volume = {560},
       number = {1},
        pages = {L83-L86},
          doi = {10.1086/324016},
archivePrefix = {arXiv},
       eprint = {astro-ph/0108472},
 primaryClass = {astro-ph},
       adsurl = {https://ui.adsabs.harvard.edu/abs/2001ApJ...560L..83M}
}

@ARTICLE{1992A&A...258..104A,
       author = {{Arenou}, F. and {Grenon}, M. and {Gomez}, A.},
        title = "{16. A tridimensional model of the galactic interstellar extinction.}",
      journal = {\aap},
         year = 1992,
        month = may,
       volume = {258},
        pages = {104-111},
       adsurl = {https://ui.adsabs.harvard.edu/abs/1992A&A...258..104A}
}

@ARTICLE{2019A&A...625A.135L,
       author = {{Lallement}, R. and {Babusiaux}, C. and {Vergely}, J.~L. and {Katz}, D. and {Arenou}, F. and {Valette}, B. and {Hottier}, C. and {Capitanio}, L.},
        title = "{Gaia-2MASS 3D maps of Galactic interstellar dust within 3 kpc}",
      journal = {\aap},
         year = 2019,
        month = may,
       volume = {625},
          eid = {A135},
        pages = {A135},
          doi = {10.1051/0004-6361/201834695},
archivePrefix = {arXiv},
       eprint = {1902.04116},
 primaryClass = {astro-ph.GA},
       adsurl = {https://ui.adsabs.harvard.edu/abs/2019A&A...625A.135L}
}

@ARTICLE{1987ARA&A..25..303C,
       author = {{Cox}, Donald P. and {Reynolds}, Ronald J.},
        title = "{The local interstellar medium.}",
      journal = {\araa},
         year = 1987,
        month = jan,
       volume = {25},
        pages = {303-344},
          doi = {10.1146/annurev.aa.25.090187.001511},
       adsurl = {https://ui.adsabs.harvard.edu/abs/1987ARA&A..25..303C}
}

@ARTICLE{2024A&A...690A.399Y,
       author = {{Yeung}, Michael C.~H. and {Ponti}, Gabriele and {Freyberg}, Michael J. and {Dennerl}, Konrad and {Liu}, Teng and {Locatelli}, Nicola and {Mayer}, Martin G.~F. and {Sanders}, Jeremy S. and {Sasaki}, Manami and {Strong}, Andy and {Zhang}, Yi and {Zheng}, Xueying and {Gatuzz}, Efrain},
        title = "{The SRG/eROSITA diffuse soft X-ray background: I. The local hot bubble in the western Galactic hemisphere}",
      journal = {\aap},
         year = 2024,
        month = oct,
       volume = {690},
          eid = {A399},
        pages = {A399},
          doi = {10.1051/0004-6361/202451045},
archivePrefix = {arXiv},
       eprint = {2410.23345},
 primaryClass = {astro-ph.GA},
       adsurl = {https://ui.adsabs.harvard.edu/abs/2024A&A...690A.399Y}
}

@ARTICLE{2000ApJS..128..171S,
       author = {{Snowden}, S.~L. and {Freyberg}, M.~J. and {Kuntz}, K.~D. and {Sanders}, W.~T.},
        title = "{A Catalog of Soft X-Ray Shadows, and More Contemplation of the 1/4 KEV Background}",
      journal = {\apjs},
         year = 2000,
        month = may,
       volume = {128},
       number = {1},
        pages = {171-212},
          doi = {10.1086/313378},
       adsurl = {https://ui.adsabs.harvard.edu/abs/2000ApJS..128..171S}
}

@ARTICLE{1990MNRAS.244..563T,
       author = {{Tenorio-Tagle}, G. and {Bodenheimer}, P. and {Franco}, J. and {Rozyczka}, M.},
        title = "{On the evolution of supernova remnants. I. Explosions inside pre-existing wind-driven bubbles.}",
      journal = {\mnras},
         year = 1990,
        month = jun,
       volume = {244},
        pages = {563},
       adsurl = {https://ui.adsabs.harvard.edu/abs/1990MNRAS.244..563T}
}

@ARTICLE{2021ApJ...920...75L,
       author = {{Linsky}, Jeffrey L. and {Redfield}, Seth},
        title = "{Could the Local Cavity be an Irregularly Shaped Str{\"o}mgren Sphere?}",
      journal = {\apj},
         year = 2021,
        month = oct,
       volume = {920},
       number = {2},
          eid = {75},
        pages = {75},
          doi = {10.3847/1538-4357/ac1feb},
       adsurl = {https://ui.adsabs.harvard.edu/abs/2021ApJ...920...75L}
}

@ARTICLE{2016Natur.532...73B,
       author = {{Breitschwerdt}, D. and {Feige}, J. and {Schulreich}, M.~M. and {Avillez}, M.~A. De. and {Dettbarn}, C. and {Fuchs}, B.},
        title = "{The locations of recent supernovae near the Sun from modelling $^{60}$Fe transport}",
      journal = {\nat},
         year = 2016,
        month = apr,
       volume = {532},
       number = {7597},
        pages = {73-76},
          doi = {10.1038/nature17424},
       adsurl = {https://ui.adsabs.harvard.edu/abs/2016Natur.532...73B}
}

@ARTICLE{2016Natur.532...69W,
       author = {{Wallner}, A. and {Feige}, J. and {Kinoshita}, N. and {Paul}, M. and {Fifield}, L.~K. and {Golser}, R. and {Honda}, M. and {Linnemann}, U. and {Matsuzaki}, H. and {Merchel}, S. and {Rugel}, G. and {Tims}, S.~G. and {Steier}, P. and {Yamagata}, T. and {Winkler}, S.~R.},
        title = "{Recent near-Earth supernovae probed by global deposition of interstellar radioactive $^{60}$Fe}",
      journal = {\nat},
         year = 2016,
        month = apr,
       volume = {532},
       number = {7597},
        pages = {69-72},
          doi = {10.1038/nature17196},
       adsurl = {https://ui.adsabs.harvard.edu/abs/2016Natur.532...69W}
}

@ARTICLE{2024ApJ...970...18L,
       author = {{Lancaster}, Lachlan and {Ostriker}, Eve C. and {Kim}, Chang-Goo and {Kim}, Jeong-Gyu and {Bryan}, Greg L.},
        title = "{Geometry, Dissipation, Cooling, and the Dynamical Evolution of Wind-blown Bubbles}",
      journal = {\apj},
         year = 2024,
        month = jul,
       volume = {970},
       number = {1},
          eid = {18},
        pages = {18},
          doi = {10.3847/1538-4357/ad47f6},
archivePrefix = {arXiv},
       eprint = {2405.02396},
 primaryClass = {astro-ph.GA},
       adsurl = {https://ui.adsabs.harvard.edu/abs/2024ApJ...970...18L}
}

@ARTICLE{2016ApJ...833..229L,
       author = {{Lee}, Eve J. and {Miville-Desch{\^e}nes}, Marc-Antoine and {Murray}, Norman W.},
        title = "{Observational Evidence of Dynamic Star Formation Rate in Milky Way Giant Molecular Clouds}",
      journal = {\apj},
         year = 2016,
        month = dec,
       volume = {833},
       number = {2},
          eid = {229},
        pages = {229},
          doi = {10.3847/1538-4357/833/2/229},
archivePrefix = {arXiv},
       eprint = {1608.05415},
 primaryClass = {astro-ph.GA},
       adsurl = {https://ui.adsabs.harvard.edu/abs/2016ApJ...833..229L}
}

@ARTICLE{2020MNRAS.493.2872C,
       author = {{Chevance}, M{\'e}lanie and {Kruijssen}, J.~M. Diederik and {Hygate}, Alexander P.~S. and {Schruba}, Andreas and {Longmore}, Steven N. and {Groves}, Brent and {Henshaw}, Jonathan D. and {Herrera}, Cinthya N. and {Hughes}, Annie and {Jeffreson}, Sarah M.~R. and {Lang}, Philipp and {Leroy}, Adam K. and {Meidt}, Sharon E. and {Pety}, J{\'e}r{\^o}me and {Razza}, Alessandro and {Rosolowsky}, Erik and {Schinnerer}, Eva and {Bigiel}, Frank and {Blanc}, Guillermo A. and {Emsellem}, Eric and {Faesi}, Christopher M. and {Glover}, Simon C.~O. and {Haydon}, Daniel T. and {Ho}, I. -Ting and {Kreckel}, Kathryn and {Lee}, Janice C. and {Liu}, Daizhong and {Querejeta}, Miguel and {Saito}, Toshiki and {Sun}, Jiayi and {Usero}, Antonio and {Utomo}, Dyas},
        title = "{The lifecycle of molecular clouds in nearby star-forming disc galaxies}",
      journal = {\mnras},
         year = 2020,
        month = apr,
       volume = {493},
       number = {2},
        pages = {2872-2909},
          doi = {10.1093/mnras/stz3525},
archivePrefix = {arXiv},
       eprint = {1911.03479},
 primaryClass = {astro-ph.GA},
       adsurl = {https://ui.adsabs.harvard.edu/abs/2020MNRAS.493.2872C}
}

@ARTICLE{2022MNRAS.512..216G,
       author = {{Grudi{\'c}}, Michael Y. and {Guszejnov}, D{\'a}vid and {Offner}, Stella S.~R. and {Rosen}, Anna L. and {Raju}, Aman N. and {Faucher-Gigu{\`e}re}, Claude-Andr{\'e} and {Hopkins}, Philip F.},
        title = "{The dynamics and outcome of star formation with jets, radiation, winds, and supernovae in concert}",
      journal = {\mnras},
         year = 2022,
        month = may,
       volume = {512},
       number = {1},
        pages = {216-232},
          doi = {10.1093/mnras/stac526},
archivePrefix = {arXiv},
       eprint = {2201.00882},
 primaryClass = {astro-ph.GA},
       adsurl = {https://ui.adsabs.harvard.edu/abs/2022MNRAS.512..216G}
}

@ARTICLE{2024MNRAS.527.6732F,
       author = {{Farias}, Juan P. and {Offner}, Stella S.~R. and {Grudi{\'c}}, Michael Y. and {Guszejnov}, D{\'a}vid and {Rosen}, Anna L.},
        title = "{Stellar populations in STARFORGE: the origin and evolution of star clusters and associations}",
      journal = {\mnras},
         year = 2024,
        month = jan,
       volume = {527},
       number = {3},
        pages = {6732-6751},
          doi = {10.1093/mnras/stad3609},
archivePrefix = {arXiv},
       eprint = {2309.11415},
 primaryClass = {astro-ph.GA},
       adsurl = {https://ui.adsabs.harvard.edu/abs/2024MNRAS.527.6732F}
}

@ARTICLE{2018A&A...619A.120K,
       author = {{Krause}, Martin G.~H. and {Burkert}, Andreas and {Diehl}, Roland and {Fierlinger}, Katharina and {Gaczkowski}, Benjamin and {Kroell}, Daniel and {Ngoumou}, Judith and {Roccatagliata}, Veronica and {Siegert}, Thomas and {Preibisch}, Thomas},
        title = "{Surround and Squash: the impact of superbubbles on the interstellar medium in Scorpius-Centaurus OB2}",
      journal = {\aap},
         year = 2018,
        month = nov,
       volume = {619},
          eid = {A120},
        pages = {A120},
          doi = {10.1051/0004-6361/201732416},
archivePrefix = {arXiv},
       eprint = {1808.04788},
 primaryClass = {astro-ph.GA},
       adsurl = {https://ui.adsabs.harvard.edu/abs/2018A&A...619A.120K}
}

@ARTICLE{2022A&A...667A.163M,
       author = {{Miret-Roig}, N. and {Galli}, P.~A.~B. and {Olivares}, J. and {Bouy}, H. and {Alves}, J. and {Barrado}, D.},
        title = "{The star formation history of Upper Scorpius and Ophiuchus. A 7D picture: positions, kinematics, and dynamical traceback ages}",
      journal = {\aap},
         year = 2022,
        month = nov,
       volume = {667},
          eid = {A163},
        pages = {A163},
          doi = {10.1051/0004-6361/202244709},
archivePrefix = {arXiv},
       eprint = {2209.12938},
 primaryClass = {astro-ph.GA},
       adsurl = {https://ui.adsabs.harvard.edu/abs/2022A&A...667A.163M}
}

@ARTICLE{2023A&A...678A..71R,
       author = {{Ratzenb{\"o}ck}, Sebastian and {Gro{\ss}schedl}, Josefa E. and {Alves}, Jo{\~a}o and {Miret-Roig}, N{\'u}ria and {Bomze}, Immanuel and {Forbes}, John and {Goodman}, Alyssa and {Hacar}, {\'A}lvaro and {Lin}, Doug and {Meingast}, Stefan and {M{\"o}ller}, Torsten and {Piecka}, Martin and {Posch}, Laura and {Rottensteiner}, Alena and {Swiggum}, Cameren and {Zucker}, Catherine},
        title = "{The star formation history of the Sco-Cen association. Coherent star formation patterns in space and time}",
      journal = {\aap},
         year = 2023,
        month = oct,
       volume = {678},
          eid = {A71},
        pages = {A71},
          doi = {10.1051/0004-6361/202346901},
archivePrefix = {arXiv},
       eprint = {2302.07853},
 primaryClass = {astro-ph.SR},
       adsurl = {https://ui.adsabs.harvard.edu/abs/2023A&A...678A..71R}
}

@ARTICLE{2024Natur.631...49S,
       author = {{Swiggum}, Cameren and {Alves}, Jo{\~a}o and {Benjamin}, Robert and {Ratzenb{\"o}ck}, Sebastian and {Miret-Roig}, N{\'u}ria and {Gro{\ss}schedl}, Josefa and {Meingast}, Stefan and {Goodman}, Alyssa and {Konietzka}, Ralf and {Zucker}, Catherine and {Hunt}, Emily L. and {Reffert}, Sabine},
        title = "{Most nearby young star clusters formed in three massive complexes}",
      journal = {\nat},
         year = 2024,
        month = jul,
       volume = {631},
       number = {8019},
        pages = {49-53},
          doi = {10.1038/s41586-024-07496-9},
archivePrefix = {arXiv},
       eprint = {2406.06510},
 primaryClass = {astro-ph.GA},
       adsurl = {https://ui.adsabs.harvard.edu/abs/2024Natur.631...49S}
}

@ARTICLE{2019MNRAS.490.1961E,
       author = {{El-Badry}, Kareem and {Ostriker}, Eve C. and {Kim}, Chang-Goo and {Quataert}, Eliot and {Weisz}, Daniel R.},
        title = "{Evolution of supernovae-driven superbubbles with conduction and cooling}",
      journal = {\mnras},
         year = 2019,
        month = dec,
       volume = {490},
       number = {2},
        pages = {1961-1990},
          doi = {10.1093/mnras/stz2773},
archivePrefix = {arXiv},
       eprint = {1902.09547},
 primaryClass = {astro-ph.GA},
       adsurl = {https://ui.adsabs.harvard.edu/abs/2019MNRAS.490.1961E}
}

@ARTICLE{2025A&A...702A..12R,
       author = {{Romano}, Leonard E.~C. and {Behrendt}, Manuel and {Burkert}, Andreas},
        title = "{SISSI: Supernovae in a stratified, shearing interstellar medium: I. The geometry of supernova remnants}",
      journal = {\aap},
         year = 2025,
        month = oct,
       volume = {702},
          eid = {A12},
        pages = {A12},
          doi = {10.1051/0004-6361/202554571},
archivePrefix = {arXiv},
       eprint = {2503.12977},
 primaryClass = {astro-ph.GA},
       adsurl = {https://ui.adsabs.harvard.edu/abs/2025A&A...702A..12R}
}

@ARTICLE{2023ApJ...947...58E,
       author = {{Ertel}, Adrienne F. and {Fry}, Brian J. and {Fields}, Brian D. and {Ellis}, John},
        title = "{Supernova Dust Evolution Probed by Deep-sea $^{60}$Fe Time History}",
      journal = {\apj},
         year = 2023,
        month = apr,
       volume = {947},
       number = {2},
          eid = {58},
        pages = {58},
          doi = {10.3847/1538-4357/acb699},
archivePrefix = {arXiv},
       eprint = {2206.06464},
 primaryClass = {astro-ph.HE},
       adsurl = {https://ui.adsabs.harvard.edu/abs/2023ApJ...947...58E}
}

@ARTICLE{2023A&A...673A.114H,
       author = {{Hunt}, Emily L. and {Reffert}, Sabine},
        title = "{Improving the open cluster census. II. An all-sky cluster catalogue with Gaia DR3}",
      journal = {\aap},
         year = 2023,
        month = may,
       volume = {673},
          eid = {A114},
        pages = {A114},
          doi = {10.1051/0004-6361/202346285},
archivePrefix = {arXiv},
       eprint = {2303.13424},
 primaryClass = {astro-ph.GA},
       adsurl = {https://ui.adsabs.harvard.edu/abs/2023A&A...673A.114H}
}

@ARTICLE{1999ApJS..123....3L,
       author = {{Leitherer}, Claus and {Schaerer}, Daniel and {Goldader}, Jeffrey D. and {Delgado}, Rosa M. Gonz{\'a}lez and {Robert}, Carmelle and {Kune}, Denis Foo and {de Mello}, Du{\'\i}lia F. and {Devost}, Daniel and {Heckman}, Timothy M.},
        title = "{Starburst99: Synthesis Models for Galaxies with Active Star Formation}",
      journal = {\apjs},
         year = 1999,
        month = jul,
       volume = {123},
       number = {1},
        pages = {3-40},
          doi = {10.1086/313233},
archivePrefix = {arXiv},
       eprint = {astro-ph/9902334},
 primaryClass = {astro-ph},
       adsurl = {https://ui.adsabs.harvard.edu/abs/1999ApJS..123....3L}
}

@ARTICLE{1957PhRv..106..620J,
       author = {{Jaynes}, E.~T.},
        title = "{Information Theory and Statistical Mechanics}",
      journal = {Physical Review},
         year = 1957,
        month = may,
       volume = {106},
       number = {4},
        pages = {620-630},
          doi = {10.1103/PhysRev.106.620},
       adsurl = {https://ui.adsabs.harvard.edu/abs/1957PhRv..106..620J}
}

@ARTICLE{2011DecA.8.206K,
       author = {{Keelin}, Thomas W. and {Powley}, Bradford W.},
        title = "{Quantile-Parameterized Distributions}",
      journal = {Decision Analysis},
         year = 2011,
        month = aug,
       volume = {8},
       number = {3},
        pages = {206-219},
          doi = {10.1287/deca.1110.0213}
}

@ARTICLE{2015ApJS..216...29B,
       author = {{Bovy}, Jo},
        title = "{galpy: A python Library for Galactic Dynamics}",
      journal = {\apjs},
         year = 2015,
        month = feb,
       volume = {216},
       number = {2},
          eid = {29},
        pages = {29},
          doi = {10.1088/0067-0049/216/2/29},
archivePrefix = {arXiv},
       eprint = {1412.3451},
 primaryClass = {astro-ph.GA},
       adsurl = {https://ui.adsabs.harvard.edu/abs/2015ApJS..216...29B}
}

@ARTICLE{2024ApJ...972..179E,
       author = {{Ertel}, Adrienne F. and {Fields}, Brian D.},
        title = "{Distances to Recent Near-Earth Supernovae from Geological and Lunar $^{60}$Fe}",
      journal = {\apj},
         year = 2024,
        month = sep,
       volume = {972},
       number = {2},
          eid = {179},
        pages = {179},
          doi = {10.3847/1538-4357/ad5a93},
archivePrefix = {arXiv},
       eprint = {2309.11604},
 primaryClass = {astro-ph.SR},
       adsurl = {https://ui.adsabs.harvard.edu/abs/2024ApJ...972..179E}
}

@ARTICLE{2021ApJ...919...35Z,
       author = {{Zucker}, Catherine and {Goodman}, Alyssa and {Alves}, Jo{\~a}o and {Bialy}, Shmuel and {Koch}, Eric W. and {Speagle}, Joshua S. and {Foley}, Michael M. and {Finkbeiner}, Douglas and {Leike}, Reimar and {En{\ss}lin}, Torsten and {Peek}, Joshua E.~G. and {Edenhofer}, Gordian},
        title = "{On the Three-dimensional Structure of Local Molecular Clouds}",
      journal = {\apj},
         year = 2021,
        month = sep,
       volume = {919},
       number = {1},
          eid = {35},
        pages = {35},
          doi = {10.3847/1538-4357/ac1f96},
archivePrefix = {arXiv},
       eprint = {2109.09765},
 primaryClass = {astro-ph.GA},
       adsurl = {https://ui.adsabs.harvard.edu/abs/2021ApJ...919...35Z}
}

@ARTICLE{2023A&A...680A..39S,
       author = {{Schulreich}, M.~M. and {Feige}, J. and {Breitschwerdt}, D.},
        title = "{Numerical studies on the link between radioisotopic signatures on Earth and the formation of the Local Bubble. II. Advanced modelling of interstellar $^{26}$Al, $^{53}$Mn, $^{60}$Fe, and $^{244}$Pu influxes as traces of past supernova activity in the solar neighbourhood}",
      journal = {\aap},
         year = 2023,
        month = dec,
       volume = {680},
          eid = {A39},
        pages = {A39},
          doi = {10.1051/0004-6361/202347532},
archivePrefix = {arXiv},
       eprint = {2309.13983},
 primaryClass = {astro-ph.SR},
       adsurl = {https://ui.adsabs.harvard.edu/abs/2023A&A...680A..39S}
}

@ARTICLE{2015ApJ...802...99K,
       author = {{Kim}, Chang-Goo and {Ostriker}, Eve C.},
        title = "{Momentum Injection by Supernovae in the Interstellar Medium}",
      journal = {\apj},
         year = 2015,
        month = apr,
       volume = {802},
       number = {2},
          eid = {99},
        pages = {99},
          doi = {10.1088/0004-637X/802/2/99},
archivePrefix = {arXiv},
       eprint = {1410.1537},
 primaryClass = {astro-ph.GA},
       adsurl = {https://ui.adsabs.harvard.edu/abs/2015ApJ...802...99K}
}

@ARTICLE{2015MNRAS.450..504M,
       author = {{Martizzi}, Davide and {Faucher-Gigu{\`e}re}, Claude-Andr{\'e} and {Quataert}, Eliot},
        title = "{Supernova feedback in an inhomogeneous interstellar medium}",
      journal = {\mnras},
         year = 2015,
        month = jun,
       volume = {450},
       number = {1},
        pages = {504-522},
          doi = {10.1093/mnras/stv562},
archivePrefix = {arXiv},
       eprint = {1409.4425},
 primaryClass = {astro-ph.GA},
       adsurl = {https://ui.adsabs.harvard.edu/abs/2015MNRAS.450..504M}
}

@ARTICLE{2018MNRAS.481.3325F,
       author = {{Fielding}, Drummond and {Quataert}, Eliot and {Martizzi}, Davide},
        title = "{Clustered supernovae drive powerful galactic winds after superbubble breakout}",
      journal = {\mnras},
         year = 2018,
        month = dec,
       volume = {481},
       number = {3},
        pages = {3325-3347},
          doi = {10.1093/mnras/sty2466},
archivePrefix = {arXiv},
       eprint = {1807.08758},
 primaryClass = {astro-ph.GA},
       adsurl = {https://ui.adsabs.harvard.edu/abs/2018MNRAS.481.3325F}
}

@ARTICLE{2017ApJ...846..133K,
       author = {{Kim}, Chang-Goo and {Ostriker}, Eve C.},
        title = "{Three-phase Interstellar Medium in Galaxies Resolving Evolution with Star Formation and Supernova Feedback (TIGRESS): Algorithms, Fiducial Model, and Convergence}",
      journal = {\apj},
         year = 2017,
        month = sep,
       volume = {846},
       number = {2},
          eid = {133},
        pages = {133},
          doi = {10.3847/1538-4357/aa8599},
archivePrefix = {arXiv},
       eprint = {1612.03918},
 primaryClass = {astro-ph.GA},
       adsurl = {https://ui.adsabs.harvard.edu/abs/2017ApJ...846..133K}
}

@ARTICLE{2012A&A...539L...1D,
       author = {{de Avillez}, M.~A. and {Breitschwerdt}, D.},
        title = "{Non-equilibrium ionization modeling of the Local Bubble. I. Tracing Civ, Nv, and Ovi ions}",
      journal = {\aap},
         year = 2012,
        month = mar,
       volume = {539},
          eid = {L1},
        pages = {L1},
          doi = {10.1051/0004-6361/201117172},
       adsurl = {https://ui.adsabs.harvard.edu/abs/2012A&A...539L...1D}
}

@ARTICLE{2019MNRAS.485.3887O,
       author = {{Ohlin}, Loke and {Renaud}, Florent and {Agertz}, Oscar},
        title = "{Supernovae feedback propagation: the role of turbulence}",
      journal = {\mnras},
         year = 2019,
        month = may,
       volume = {485},
       number = {3},
        pages = {3887-3894},
          doi = {10.1093/mnras/stz705},
archivePrefix = {arXiv},
       eprint = {1902.00028},
 primaryClass = {astro-ph.GA},
       adsurl = {https://ui.adsabs.harvard.edu/abs/2019MNRAS.485.3887O}
}

@ARTICLE{2025A&A...701L...5R,
       author = {{Romano}, Leonard E.~C. and {Owen}, Ellis R. and {Nagamine}, Kentaro},
        title = "{Starburst-driven galactic outflows: Unveiling the suppressive role of cosmic ray halos}",
      journal = {\aap},
         year = 2025,
        month = sep,
       volume = {701},
          eid = {L5},
        pages = {L5},
          doi = {10.1051/0004-6361/202554590},
archivePrefix = {arXiv},
       eprint = {2503.13261},
 primaryClass = {astro-ph.GA},
       adsurl = {https://ui.adsabs.harvard.edu/abs/2025A&A...701L...5R}
}

@ARTICLE{2006MNRAS.373..993F,
       author = {{Fuchs}, B. and {Breitschwerdt}, D. and {de Avillez}, M.~A. and {Dettbarn}, C. and {Flynn}, C.},
        title = "{The search for the origin of the Local Bubble redivivus}",
      journal = {\mnras},
         year = 2006,
        month = dec,
       volume = {373},
       number = {3},
        pages = {993-1003},
          doi = {10.1111/j.1365-2966.2006.11044.x},
archivePrefix = {arXiv},
       eprint = {astro-ph/0609227},
 primaryClass = {astro-ph},
       adsurl = {https://ui.adsabs.harvard.edu/abs/2006MNRAS.373..993F}
}

@INCOLLECTION{1998LNP...506..483K,
       author = {{Kahn}, F.~D.},
        title = "{The Galactic Fountain}",
    booktitle = {IAU Colloquium 166: The Local Bubble and Beyond},
         year = 1998,
       editor = {{Breitschwerdt}, Dieter and {Freyberg}, Michael J. and {Truemper}, Joachim},
       volume = {506},
        pages = {483-494},
          doi = {10.1007/BFb0104770},
       adsurl = {https://ui.adsabs.harvard.edu/abs/1998LNP...506..483K}
}

\appendix

\section{The SISSI simulations} \label{app:SISSI}

The SISSI simulations are a suite of highly resolved ($\Delta x \lesssim 1\, \text{pc}$) hydrodynamical simulations of SNe exploding in thirty distinct regions of the multi-phase ISM of an isolated Milky-Way-like galaxy. The ISM of the SISSI galaxy is shaped by the complex interplay of stellar feedback, realistic radiative cooling and heating (M. Behrendt et al. in prep.), secular evolution in an axisymmetric background gravitational potential, and local gravitational collapse and star formation. 

As such the SISSI simulations represent a natural next step in terms of physical realism for models of SNRs in environments of increasing realism and complexity \citep[e.g.][]{2015ApJ...802...99K, 2018MNRAS.481.3325F, 2017ApJ...846..133K}. Due to the large range of probed physical conditions they complement simulations of specific SNRs such as the LB \citep[e.g.][]{2006A&A...452L...1B, 2012A&A...539L...1D, 2023A&A...680A..39S} by drawing general conclusions about SNR evolution in a realistic ISM. 

In SISSI we considered three separate explosion scenarios: (i) N1 corresponds to a single SN explosion at each site at the beginning of the zoom-in period (t=0), (ii) N10 corresponds to 10 simultaneous SN explosions at each site at t=0, and (iii) N1x10 corresponds to a single SN at each site, every 1 Myr starting from t=0. 
The zoom-in period begins after a prolonged ISM-generation period at lower resolution, where stellar feedback is modeled in an effective way. During the zoom-in period lasting 10 Myr, the effective stellar feedback is shut-off in order to not interfere with the zoomed-in SNe.

Some key features of the SISSI simulations include
\begin{enumerate}
    \item Ambient densities in the range $10^{-3} \lesssim n_{\text{H}} \left[\text{cm}^{-3}\right] \lesssim 10^{3}$.
    \item A multi-phase ISM with a typical gas velocity dispersion of $\sim 10\,\text{km/s}$ with significant scatter.
    \item Explosion sites at three distinct galactocentric radii ($R \sim 2, \, 4.5, \, \text{and} \, 8 \, \text{kpc}$, corresponding to orbital timescales of $\sim 50 - 200\, \text{Myr}$.
\end{enumerate}

Here we recapitulate some of the key differences between SNRs exploding in the ISM of the SISSI galaxy versus SNRs exploding in uniform media, found in \citetalias{2025A&A...702A..12R}:
\begin{enumerate}
    \item Strong deviations from spherical symmetry (Minor-to-major-ratios $\lesssim 0.2$, semi-major-to-major-ratios $\lesssim 0.4$).
    \item A systematically larger size.
\end{enumerate}
These differences generally manifest after a few Myr or equivalently a few percent of an orbit, likely due galactic shear, the galactic density structure and the coupling to pre-existing blow-out regions. At the low number of SNe considered in SISSI, SBs are not expected to break out of the disk significantly, unless they expand through already carved out low-density channels \citep{2023A&A...680A..39S, 2025A&A...701L...5R}.

The parameter space covers the physical conditions of the LB (e.g., ambient density, geometry), allowing us to draw tentative conclusions about its origin, under the assumption that it is not too different from a typical SNR in a similar environment from the SISSI simulation suite.  

\subsection{Comparison to simulations of the local ISM}

There are a number of studies studying the properties and history of the Local Bubble, using dedicated simulations of SNe in the local ISM \citep{2006A&A...452L...1B, 2012A&A...539L...1D, 2023A&A...680A..39S}. In these studies, a stratified $\sim 1\, \text{kpc} \times 1\, \text{kpc}$ ISM patch, resolved with a maximum resolution of $\sim 1 \, \text{pc}$, and physical conditions roughly matching those in the local ISM is set up before the stellar feedback input (SNe and in more recent studies stellar winds) from a sample of nearby stellar groups, associated with the present-day Sco-Cen association are allowed to perturb the medium and carve out a SB identified with the LB.

These studies are in excellent agreement with various observables in the local ISM, such as the column-density of \ion{O}{VI} and other ions \citep{2012A&A...539L...1D} and the deposition of radio-isotopes in the solar system \citep{2016Natur.532...73B}. These simulations suggest an age of the LB of 10-15 Myr since the onset of SN-explosions in Sco-Cen.

Despite these successes it should be noted, that several simplifications have been made that might affect their results.
\begin{enumerate}
    \item The simulations neglect the self-gravity of the gas, which might affect the structure of the ISM, the orbits of the stellar-groups depositing their feedback and the expansion of the LB once it has sufficiently slowed down, especially in those regions approaching denser regions of the local ISM. 
    \item Galactic rotation and radial gradients are neglected possibly affecting the geometry of the stars and gas of the ISM into which the simulated LB expands. At the suggested age of $\sim 14 \, \text{Myr}$ these effects will likely already be relevant as shown in \citetalias{2025A&A...702A..12R}.
    \item A slightly higher average density and in \citet{2023A&A...680A..39S} a uniform midplane density field likely affect the typical size of the simulated LB in these simulations. The absence of a medium, structured by turbulence and self-gravity, leads to the absence of a volume-filling warm low-density phase and low-density channels through which SNRs can expand more rapidly than in a uniform medium \citep{2019MNRAS.485.3887O}.
    \item There are significant uncertainties in the ages and explosion times of the progenitor clusters of the LB \citep{2023A&A...673A.114H, 2023A&A...678A..71R, 2024Natur.631...49S}. By drawing only a single sample these studies neglect these studies likely underestimate the uncertainty arising from this simplification.
\end{enumerate}

In SISSI these simplifications are avoided by placing the SNe in the self-consistently generated ISM of an entire Milky-Way-like galaxy, at the cost of the anyway quite uncertain spatial and temporal sequence of SN events in the local ISM.

\section{Analysis of the LB using 3D dust maps} \label{app:local_bubble}

\begin{table*}[t]
\caption{\centering Properties of the Local Bubble}
\label{tab:local_bubble}
\begin{center}
\resizebox{0.7\textwidth}{!}{
\begin{tabular}{l|l l l|l}
  Property & This work  & \citetalias{2022Natur.601..334Z} & \citetalias{2024ApJ...973..136O} & Unit\\\hline
  Effective Radius & $212.3\pm1.0_{\text{stat}}$  & $165\pm6$ & 170\tablefootmark{a} & pc\\
  Mass & $6.2 \pm 0.7_{\text{sys}} \pm 0.07_{\text{stat}}$  & $14^{+6.5}_{-6.2}$ & $6.0 \pm0.7$\tablefootmark{b} & $10^5 \, \text{M}_{\odot}$\\
  Hydrogen number density &  $0.50 \pm 0.06_{\text{sys}} \pm 0.006_{\text{stat}}$ & $2.71^{+1.57}_{-1.02}$ & 0.61\tablefootmark{c} & cm$^{-3}$\\
  Momentum & $\left(3.0 \pm 1.7_{\text{sys}}\pm 0.05_{\text{stat}} \right)$\tablefootmark{d} $\times \,(\text{Myr} \,/ \,t_{\text{age}})$ & $1.0^{+0.4}_{-0.4}$ & -- & $10^7 \, \text{M}_{\odot} \, \text{km s}^{-1}$\\
  \hline
  Minor-to-major-ratio & $0.469 \pm 0.007$ & -- & -- & -- \\
  Semi-major-to-major-ratio & $0.562 \pm 0.011$ & -- & -- & -- \\
  Pitch angle (major axis) & $-19.0 \pm 1.0$ & -- & -- & $^{\circ}$ \\
  Polar direction (major axis) & $0.665 \pm 0.011$ & -- & -- & -- \\
  Pitch angle (minor axis) & $-66.1 \pm 2.7$ & -- & -- & $^{\circ}$ \\
  Polar direction (minor axis) & $0.607 \pm 0.022$ & -- & -- & -- \\\hline
\end{tabular}
}
\end{center}
\begin{center}
\resizebox{0.7\textwidth}{!}{
\tablefoottext{a}{Median peak distance}
\tablefoottext{b}{Mass enclosed within shell}
\tablefoottext{c}{Median peak density}
\tablefoottext{d}{Bias corrected (see Appendix~\ref{app:momentum})}
}
\end{center}
\end{table*}

We recovered the properties of the LB by applying a similar analysis as \citetalias{2024ApJ...973..136O}, based on the 3D dust maps of \citetalias{2024A&A...685A..82E}, to obtain estimates for the LB's geometry, through the means of the shape tensor, defined in \citetalias{2025A&A...702A..12R} (See also Appendix~\ref{app:shape_tensor} for the definition).
Some steps of our analysis required the original data products of \citetalias{2024A&A...685A..82E}, so we opted for an independent analysis, instead of directly using the data products of \citetalias{2024ApJ...973..136O}.

We followed the same basic steps of smoothing, peak finding and finally an analysis of the mass distribution enclosed by the peaks, used by \citetalias{2024ApJ...973..136O}, with only minor differences.
In contrast to \citetalias{2024ApJ...973..136O} we directly worked with the logarithmically spaced grid of the 12 sample dust maps rather than sampling the 3D dust maps on a linearly spaced grid of the mean of the sample maps. 
Moreover, we weighted the smoothing kernel to explicitly account for the non-uniform volume-elements $\text{d}V\propto r^2 \,\text{d}r$  (Appendix~\ref{app:smoothing-kernel}) and we linearly extrapolated the density field beyond the grid boundaries to better capture the profile-shape in their vicinity (Appendix~\ref{app:boundaries}).
In Appendix~\ref{app:evaluation} we show that our results largely agree with those of \citetalias{2024ApJ...973..136O}, demonstrating the robustness of the method.
The following steps of peak identification and conversion between differential extinction and hydrogen number density are identical to \citetalias{2024ApJ...973..136O}, who estimate a systematic uncertainty of $\sim 10\, \%$ for the conversion factor, which we propagate into our mass and momentum estimates.

To obtain the average density of the LB, we divided the total mass, corresponding to the volume integral of the density from $r = 0$ to the outer edge of the shell, by the volume.
The outer edge is defined as the first point past the peak with half the peak's prominence. 
We accounted for the mass contained within the central $\sim 69 \, \text{pc}$, where the differential extinction is missing from the dust maps, by assuming a constant hydrogen number density along each line-of-sight within this central volume,
proportional to the integrated extinction at $\sim 69 \, \text{pc}$. The mass in this region accounts for $\sim 7 \times 10^3 \, \text{M}_{\odot}$, providing a conservative upper limit on the systematic uncertainty due to this assumption.

We estimated the momentum as a function of age by assuming a rapidly cooling wind expansion, $R \propto t^{1/2}$ \citep{2022ApJS..262....9O, 2024ApJ...970...18L}, and homologous expansion $v \propto r$, giving
\begin{equation} \label{eq:momentum}
    p_{\text{LB}}\left(t_{\text{age}}\right) = \int_{\text{LB}} v \, \text{d}M \approx \frac{1}{2\,t_{\text{age}}}\int\int_0^{R_{\text{out}}\left(\Omega\right)}  \mu n_{\text{H}}\left(r, \Omega\right) r^3 \text{d}r\,\text{d}\Omega ~,
\end{equation}
where radial integrals sum over radial bins of the unsmoothed profile within the outer shell radius, angular integrals sum over all \textsc{Healpix} lines-of-sight with constant $\text{d}\Omega \approx 1.6 \times 10^{-5}\,\text{rad}^2$, and we used $v \approx r/\left(2t_{\text{age}}\right)$, where $t_{\text{age}}$ is the age of the LB. 
As shown in Appendix~\ref{app:momentum}, Eq.~\ref{eq:momentum} overestimates the momentum by a factor of $\sim 2$, which we accordingly corrected for. 
The resulting momentum as a function of time for fixed size $R$ is consistent with SB expansion under a constant momentum injection rate\footnote{Constant momentum injection does not necessarily imply momentum-conservation. For instance in the rapidly cooling regime, the momentum injected into the shell is boosted by the thermal energy of the bubble.} $\dot{p}_{\text{SN}}$, giving $p_\text{LB} \propto R^4 / t \propto \dot{p}_{\text{SN}} \, t$.

The velocity estimate for momentum-driven expansion provides a conservative lower limit for the momentum. In the case of purely energy-driven expansion, the pre-factor for the velocity would be $3/5$, and lead to a $20 \, \%$ higher momentum estimate. The LB might operate in between these regimes.

The assumption of homologous expansion leads to further systematic uncertainties, which we estimate by considering the momentum in the case that all of the gas along each line-of-sight is expanding at the same speed. In this approximation, the momentum is $\sim 12 \, \%$ higher than our estimate from homologous expansion. Conversely a more concentrated velocity distribution within the LB will lead to a slightly lower momentum. We thus estimate the uncertainty due to the assumed velocity structure to be on the order of $\sim 10 \, \%$, though this is likely highly correlated with the factor-of-two correction factor that we employed.

\citetalias{2024ApJ...973..136O} have identified an open ``chimney''-like topology towards the nothern galactic pole, suggesting that it might be part of a larger blow-out structure that is not covered by the 3D dust maps. While our shell-identification method recovers a well-defined shell position for each line-of-sight, the possibility of  such a chimney is not entirely ruled out, since the 3D dust maps are increasingly uncertain towards low-density regions such as the galactic poles. Thus in principle, our derived estimates for momentum, mass and size are to be understood as lower limits.

We summarize the thus obtained properties of the LB in Tab. \ref{tab:local_bubble}. 
For comparison, we also show the values obtained by \citetalias{2022Natur.601..334Z} and \citetalias{2024ApJ...973..136O} where available. 
Overall our results are similar to the more recent investigation of \citetalias{2024ApJ...973..136O}, however we also calculated the momentum, which we can use to estimate the age of the LB. 
Uncertainties are dominated by systematic uncertainties at the 60 percent level.

\subsection{Modified smoothing Kernel}\label{app:smoothing-kernel}

\citetalias{2024ApJ...973..136O} used a Gaussian smoothing-kernel $W$ to obtain smooth density profiles before applying their peak-finding algorithm. This procedure is not manifestly mass-conserving and might lead to numerical artifacts that can bias the results.

In order to account for this potential bias, we used an explicitly mass-conserving approach, by introducing weights that account for the non-uniformity of the volume elements $\text{d}V = r^2 \, \text{d}r \, \text{d}\Omega$.
In particular, the condition that the mass is identical between the unsmoothed and smoothed profiles can be written as
\begin{equation}\label{eq:mass_conservation}
    M = \sum_{i} \text{d}V_{i} \, \rho_i = \sum_{i} \text{d}V_{i} \, \bar{\rho}_{i} ~,
\end{equation}
where the smoothed density
\begin{equation}
    \bar{\rho}_{i} = \sum_{j} W_{ij}\rho_j ~.
\end{equation}
Condition \ref{eq:mass_conservation}, leads to the normalization condition for the smoothing matrix $W$
\begin{equation}
    \text{d}V_{j} = \sum_{i} \text{d}V_{i} W_{ij} ~.
\end{equation}
We find that with these modifications a smoothing length of $\sigma_{\text{smth}} = 9 \, \text{pc}$, slightly above the value used by \citetalias{2024ApJ...973..136O} yields robust results.
However, we find that unless one uses the smoothed density profile for integrals -- which like \citetalias{2024ApJ...973..136O}, we do not -- the differences due to the mass-conserving Gaussian kernel are only minor, reinforcing the robustness of both methods.

\subsection{The role of the boundary treatment}\label{app:boundaries}

\begin{figure}
 \includegraphics[width=\linewidth, clip=true]{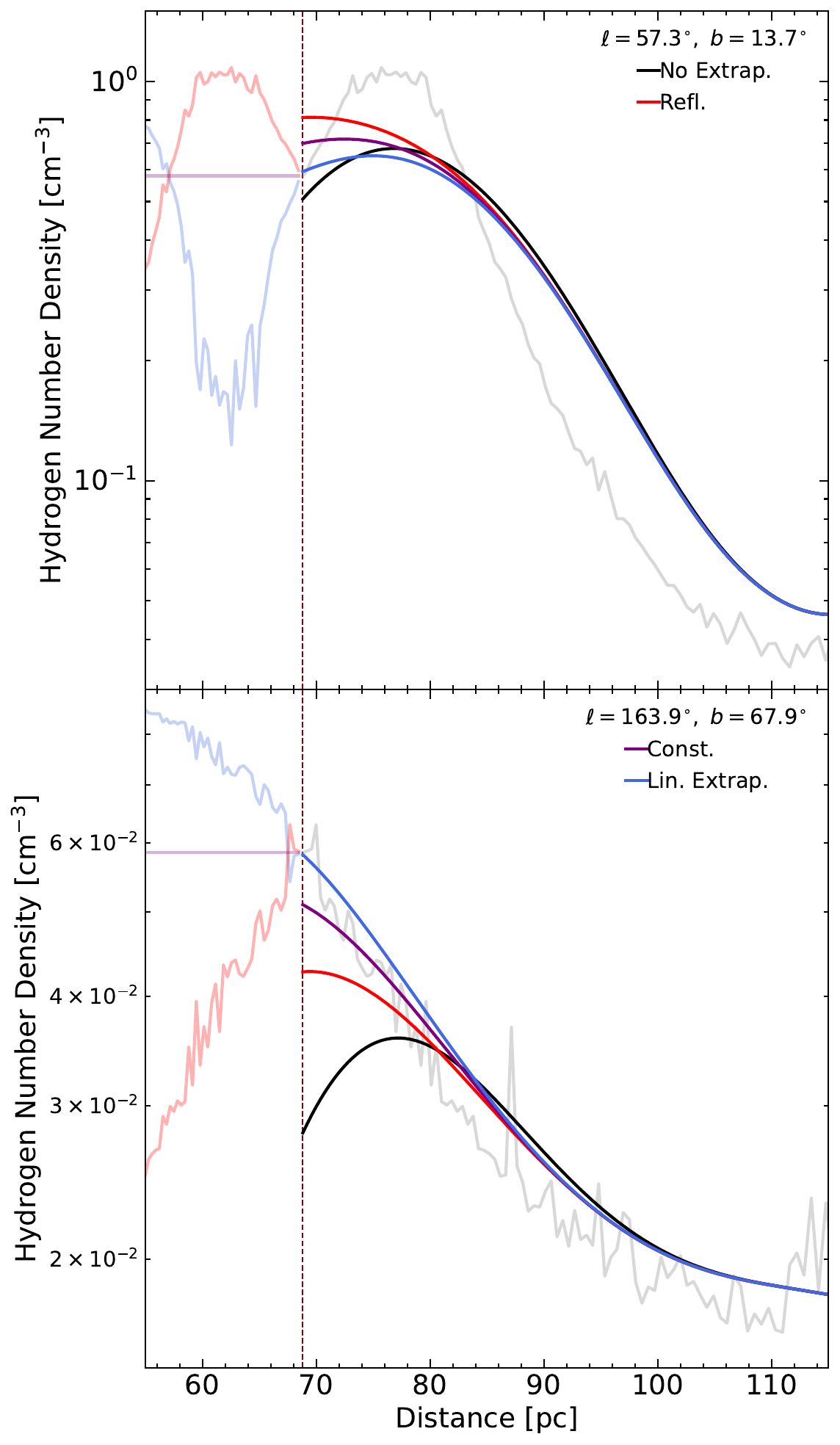}
 \caption{Smoothed density profiles using different boundary treatments for two characteristic lines-of-sight.
 The boundary is indicated by the dashed vertical line.
 The original data is shown as a gray line, and the extrapolations beyond the boundary are shown at decreased opacity.
 The top panel shows a line-of-sight with a peak near the inner boundary, while the bottom panel shows a line of sight with a steep decline near the boundary.
 With reflective boundary conditions, the peak is missed entirely, while simple normalization adds a spurious peak even in the case of the steep decline.
  } 
 \label{fig:boundaries}
\end{figure}

The choice of the boundary conditions can qualitatively affect the results, by introducing spurious peaks or erasing physical peaks near the boundary.

The simplest way to deal with the boundaries is not to deal with them at all.
In particular, consider the smoothing matrix $W$ and assume that it is properly normalized:
\begin{equation}
    \sum_{i} w_{i}W_{ij} = w_{j} ~,
\end{equation}
where $w_{i}$ are some weights (e.g., volume, a constant, etc.), then weights of the points near the boundary are systematically boosted.
This results in smoothed data that will systematically lie below (above) the original data values, if they exhibit a negative (positive) gradient near the boundary.
On the other hand, if proper normalization is omitted, points near the boundaries lose weight and the smoothed profile is always systematically lower than the data. 

A similar way to deal with boundaries are so-called reflective boundary conditions, where the values of the data as well as their positions are mirrored beyond the boundary, namely, $x_{-i} = x_0 - (x_i - x_0)$ and $y_{-i} = y_{i}$. 
This method increases the weight of points near the boundary, but in a more controlled way.

 Another common method is to set the data outside of the boundary region to a constant value, for instance the value of the data at the boundary.
 This leads to a slight over- (under-) estimation near the boundary, in the case of a positive (negative) slope in the original data.

 Finally, linear extrapolation of the data beyond the boundary promises to capture steep gradients near the boundaries well.
In particular, here we consider a linear extrapolation centered around the boundary point:
\begin{equation}
    y_{-i} = y_0 + \frac{y_{i} - y_{0}}{x_{i} - x_{0}} \left(x_{-i} - x_{0}\right) ~.
\end{equation}

Since all of these methods are linear, they can be directly incorporated into the smoothing kernel matrix $W$, which can be precomputed.

In Fig. \ref{fig:boundaries} we show how these various methods compare for two directions, exhibiting some of the features where differences between the methods are likely to arise, namely, a peak near the boundary, and a steep negative gradient.

In the case of the peak near the boundary it is critical, that the smoother does not remove the peak. 
However, we find that this is the case for reflective boundary conditions.
In the other cases with extrapolation, a shallow peak remains, indicating that in certain cases the peak might still disappear.
Only in the case without extrapolation, we recover a strong peak.
On the other hand, in the case of a steep negative gradient, it is important that no spurious peaks emerge, a condition that is not satisfied for the smoother without extrapolation alone.

In summary, reflective boundary conditions as well as no extrapolation are not well suited for the task at hand, while linear extrapolation and constant boundaries appear to capture the most important features.
Due to the slightly better performance at capturing peaks near the boundary, we opted for linear extrapolation.

\subsection{Comparison to previous work}\label{app:evaluation}

\begin{table}[t]
\caption{Comparison of LB properties derived using different methods.}
\label{tab:comparison}
\resizebox{\linewidth}{!}{
\begin{tabular}{l|c c c|l}
  Property & This work  & 
 Repr.
  & O'Neill+ (2024) & Unit\\
  \hline
  Inner Edge & $151^{+177}_{-71}$ & $151^{+179}_{-69}$ & $150^{+172}_{-70}$ & pc\\
  Peak Distance & $172^{+191}_{-80}$ & $169^{+192}_{-77}$ & $170^{+192}_{-79}$ & pc\\
  Outer Edge & $194^{+206}_{-89}$ & $190^{+201}_{-89}$ & $191^{+204}_{-90}$ & pc\\
  Shell Thickness & $38^{+87}_{-24}$ & $34^{+80}_{-22}$& $35^{+88}_{-22}$ & pc\\
  Shell Mass & $6.5$ & $5.8$ & $6.0 \pm 0.7$& $10^5\,\text{M}_{\odot}$\\
  \hline      
\end{tabular}
}
\end{table}

In order to ensure that our results are robust, we tried to reproduce the shell properties reported by \citetalias{2024ApJ...973..136O} using the method described in their paper as well as the method described in Sec. \ref{app:local_bubble}, both applied to the sample mean 3D dust map.

In particular, we sampled the dust map between 69 and 1244 pc at uniform 1 pc intervals and smoothed it with a Gaussian kernel with a smoothing length of $\sigma_{\text{smth}} = 7 \,\text{pc}$ with reflective boundary conditions.
Peak finding and the location of the inner and outer peak boundaries are done identically between the two methods.

In Tab. \ref{tab:comparison} we compare the thus obtained results.
For the distances to the peak (boundaries) and the width of the shell we show the median, including $\pm2\sigma$ uncertainties, corresponding to the $2.28^{\text{th}}$ and $97.72^{\text{th}}$ percentiles, while for the shell mass, namely, the mass between the inner and outer edge of the shell, we simply show the value obtained from the analysis of the sample mean 3D dust map.
While we could not reproduce the results of \citetalias{2024ApJ...973..136O} exactly, our results are reasonably close and small differences are likely due to minor details omitted in their description of their analysis.

Curiously, in the reconstruction of the shell by \citetalias{2024ApJ...973..136O} there are small holes in a few directions where their method cannot identify any prominent peaks, while our method fails to reproduce them. 
We inspected the profiles and peak identification for a handful of these directions and do indeed see prominent peaks, indicating that these holes might have arisen from the initial sampling step. Due to their small number, they do not meaningfully affect our conclusion about the average geometry.

\section{Evaluation of analysis methods}

\subsection{Momentum estimate} \label{app:momentum}

\begin{figure}
 \includegraphics[width=\linewidth, clip=true]{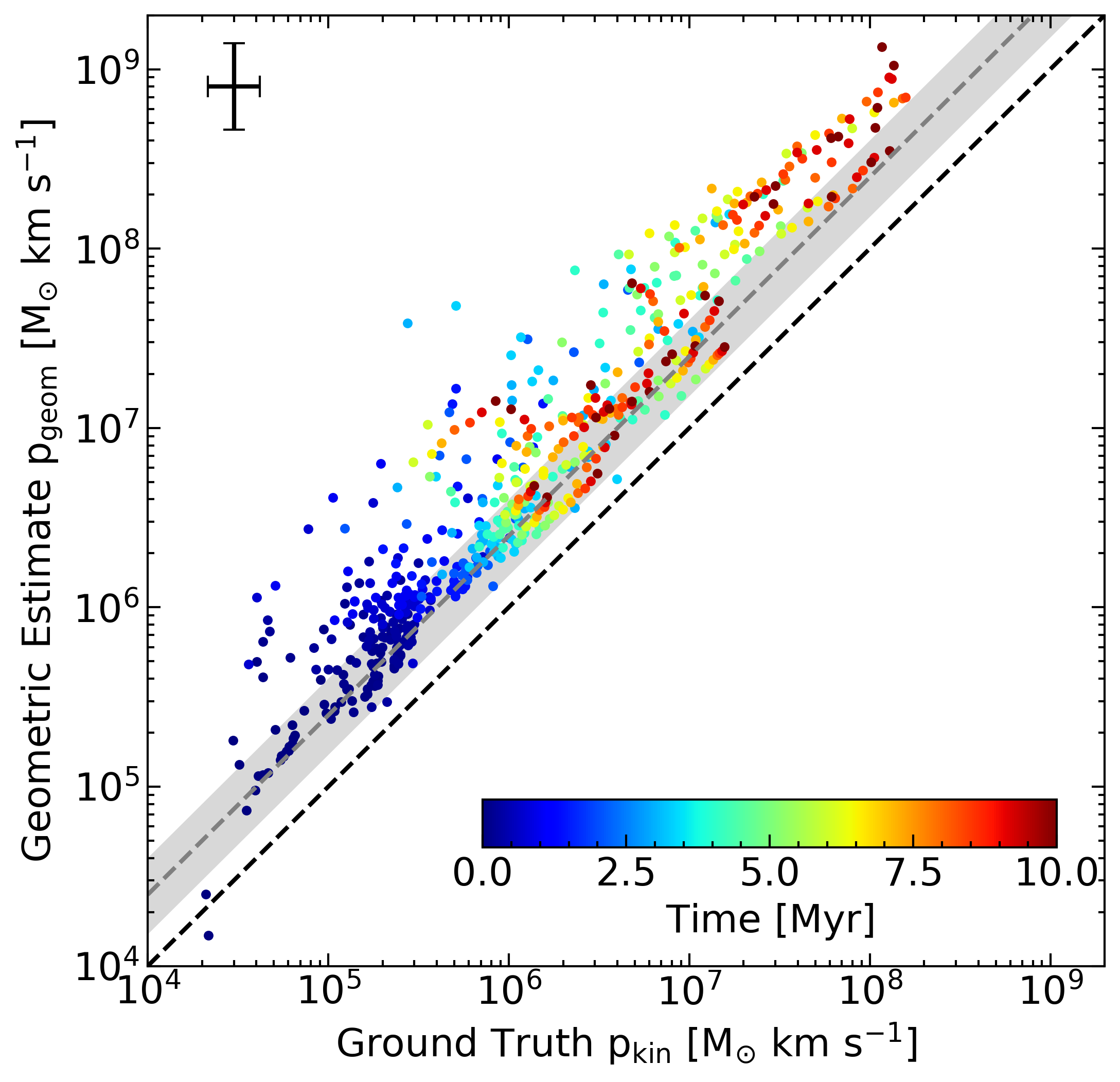}
 \caption{Comparison of the geometric momentum estimate Eq. \ref{eq:momentum} to the true momentum for the simulated sample of SBs from the SISSI simulation at different points in time, as indicated by the color scheme. A representative error bar corresponding to uncertainties due to different ways of defining the center (geometric center vs. center of mass) of the SBs and their volume (threshold value for passive-scalar tracer-variable) is shown in the upper-left corner.
 The geometrical momentum estimate is slightly biased towards larger values, by a factor of $\sim 2$.} 
 \label{fig:momentum}
\end{figure}

In order to assess the accuracy of the momentum estimate in Eq. \ref{eq:momentum}, we computed such geometrical momentum estimates for the SISSI sample of simulated SNRs at different points in time and compared them to the true momentum obtained from the velocity field. 
The result of this comparison is shown in Fig. \ref{fig:momentum}, which shows that the geometrical momentum estimate is slightly biased towards greater values, by a factor of $\sim 2$, indicative of the (expected) departure from homologous expansion.

The choice of the reference central point, as well as other geometric effects, for instance due to the definition of the SNRs volume may affect this estimate and introduce additional systematic uncertainties reflected by the representative error bar shown in Fig. \ref{fig:momentum}.
The error bar, corresponding to a range of extreme scenarios provides a conservative upper limit for the systematic uncertainty induces by these effects of $\sim 0.24 \, \text{dex}$.

\subsection{Number of SN explosions} \label{app:N_SN}

\begin{figure}
 \centering
 \includegraphics[width=\linewidth, clip=true]{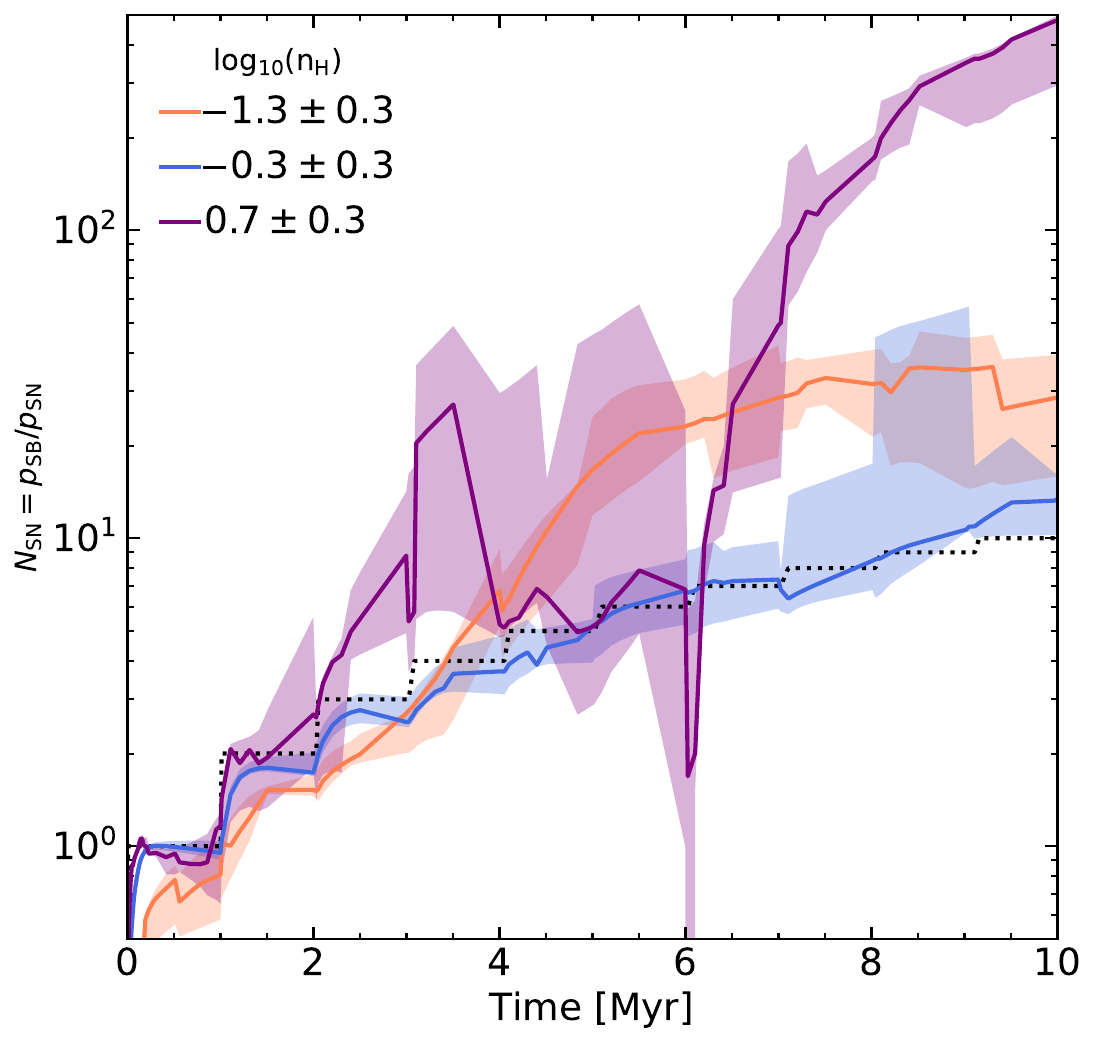}
 \caption{Time-evolution of the SN count estimate Eq. \ref{eq:N_SN} for the simulated sample of SBs in different density ranges. The true SN count is shown as a black dashed line.
 At extremely high and low ambient densities the estimate appears to overestimate the number of SNe, while for moderate densities, around the density of the LB, the estimate is quite accurate, within about $\lesssim 50\%$.
  } 
 \label{fig:SN_count}
\end{figure}

We assessed the accuracy of the SN count estimate in Eq. \ref{eq:N_SN},
using our sample of simulated SBs, by comparing the thus estimated SN count with the actual number of injected SNe.
The results are shown in Fig. \ref{fig:SN_count} which suggests that for moderate densities comparable to that of the LB, the estimate is fairly accurate, while for higher and lower densities it increasingly overestimates the number of SNe as time passes.

This overestimate might be driven by mergers of the simulated, controlled SBs with preexisting SBs in their environment, for which we have no handle to accurately count the number of SNe that they would contribute.
At low densities, SBs grow much larger and are thus more likely to merge with other large SBs, while at high densities, the star-formation rate is elevated, leading to a more active environment, with a high density of SNRs.

\subsection{Shape Tensor}\label{app:shape_tensor}

In \citetalias{2025A&A...702A..12R} the shape tensor is defined as
\begin{equation}
    S_{ij} = V_{\text{SNR}}^{-1}\int_{\text{SNR}} \left(\left\lVert \mathbf{x}\right\rVert^2\delta_{ij} - x_{i}x_{j}\right)\text{d}^3\mathbf{x} ~.
\end{equation}
By assuming an approximately ellipsoidal shape, the three ellipsoidal radii, are defined by
\begin{equation}
    r_{i} = \sqrt{2.5\left(\text{tr}\left(S\right) - 2S_{i}\right)} ~,
\end{equation}
where $S_{i}$ are the eigenvalues of $S_{ij}$ and $\text{tr}\left(S\right)$ is the trace.
The smallest, intermediate and largest eigenvalues correspond the minor $a$, semi-major $b$ and major $c$ axis, respectively.
The effective size of an SNR is the geometric mean of the three eigenvalues
\begin{equation}
    r_{\text{eff}} = \left(a b c\right)^{1/3}~.
\end{equation}

To determine the alignment of the LB within the Galaxy, we measured the pitch angle $\alpha$ and polar direction $\text{cos}\left(\theta\right)$ for both the major and minor axes.
The pitch angle is defined relative to the direction of Galactic rotation, with $\alpha = 90^\circ$ and $\alpha = -90^\circ$ corresponding to the Galactic center and anti-center, respectively.
The magnitude of the polar direction is 0 (1) for directions parallel (perpendicular) to the Galactic plane. 

\section{Models for radiative blastwaves in uniform media} \label{app:SB_models}

We distinguish between blastwave models driven by a single explosion at $t=0$ and blastwaves that are driven by continuous energy- and momentum-injection.

The dynamics of a radiative blastwave driven by a single explosion are determined the conservation of the radial momentum that was acquired before the onset of radiative cooling. This phase is also known as the momentum-conserving snowplow phase \citep[e.g.,][]{2022ApJS..262....9O}. 
The shock radius of the momentum-conserving snowplow as a function of expansion time $t$ is given by \citep[e.g.,][]{2022ApJS..262....9O}
\begin{equation}\label{eq:MCS}
    R_{\text{MCS}} = \left(\frac{3 N_{\text{SN}} \, \hat{p}_{\text{SN}} \, t}{\pi \, \rho_{\text{ISM}}}\right)^{1/4} ~,
\end{equation}
where $N_{\text{SN}}$ is the number of SNe exploding at $t=0$, $\hat{p}_{\text{SN}}$ is the momentum injected per SN, given by Eq. \ref{eq:momentum_per_SN} and $\rho_{\text{ISM}}$ is the ambient density.

The dynamics of continuously driven blastwaves are more uncertain and seem to depend on the detailed balance of cooling and energy injection \citep{2019MNRAS.490.1961E, 2022ApJS..262....9O, 2024ApJ...970...18L}.
If the cooling is relatively gentle and a large fraction of the injected energy can reach the radiative shell, the dynamics of the blastwave match those of an energy-driven wind with a scaled down energy injection rate \citep{2019MNRAS.490.1961E}
\begin{equation}
    R_{\text{EDW}} = \xi_{\text{EDW}} \left(\frac{\left(1-\theta\right) \,  E_{\text{SN}} \, t^3}{\Delta t_{\text{SN}} \, \rho_{\text{ISM}}}\right)^{1/5} ~,
\end{equation}
where $\xi_{W} \sim 0.88$ is determined by considering the internal structure of the blastwave, $E_{\text{SN}} = 10^{51}\, \text{erg}$ is the total injected per SN,  
$\left(1-\theta\right) \sim 0.3$\ \citep{2022Natur.601..334Z} is a reduction factor, accounting for the effect of radiative cooling and $\Delta t_{\text{SN}}$ is the average time between SN explosions.
On the other hand, if cooling is sufficiently rapid each SN contributes the same momentum $\hat{p}_{\text{SN}}$ and the dynamics are determined by the conservation of the momentum, which is imparted onto the radiative shell at a constant rate \citep{2022ApJS..262....9O, 2024ApJ...970...18L}
\begin{equation}
    R_{\text{MDW}} = \left(\frac{3  \hat{p}_{\text{SN}} \, t^2}{2\pi \, \Delta t_{\text{SN}} \, \rho_{\text{ISM}}}\right)^{1/4} ~,
\end{equation}
which may be compared to Eq. \ref{eq:MCS} upon substituting $N_{\text{SN}}\left(t\right) = t \, / \, \Delta t_{\text{SN}}$. This model is often referred to as momentum-driven or rapidly-cooling wind \citep{2024ApJ...970...18L}.

\section{Passage of the solar system} \label{app:solar_system}
\begin{figure}
\centering
 \includegraphics[width=\linewidth, clip=true]{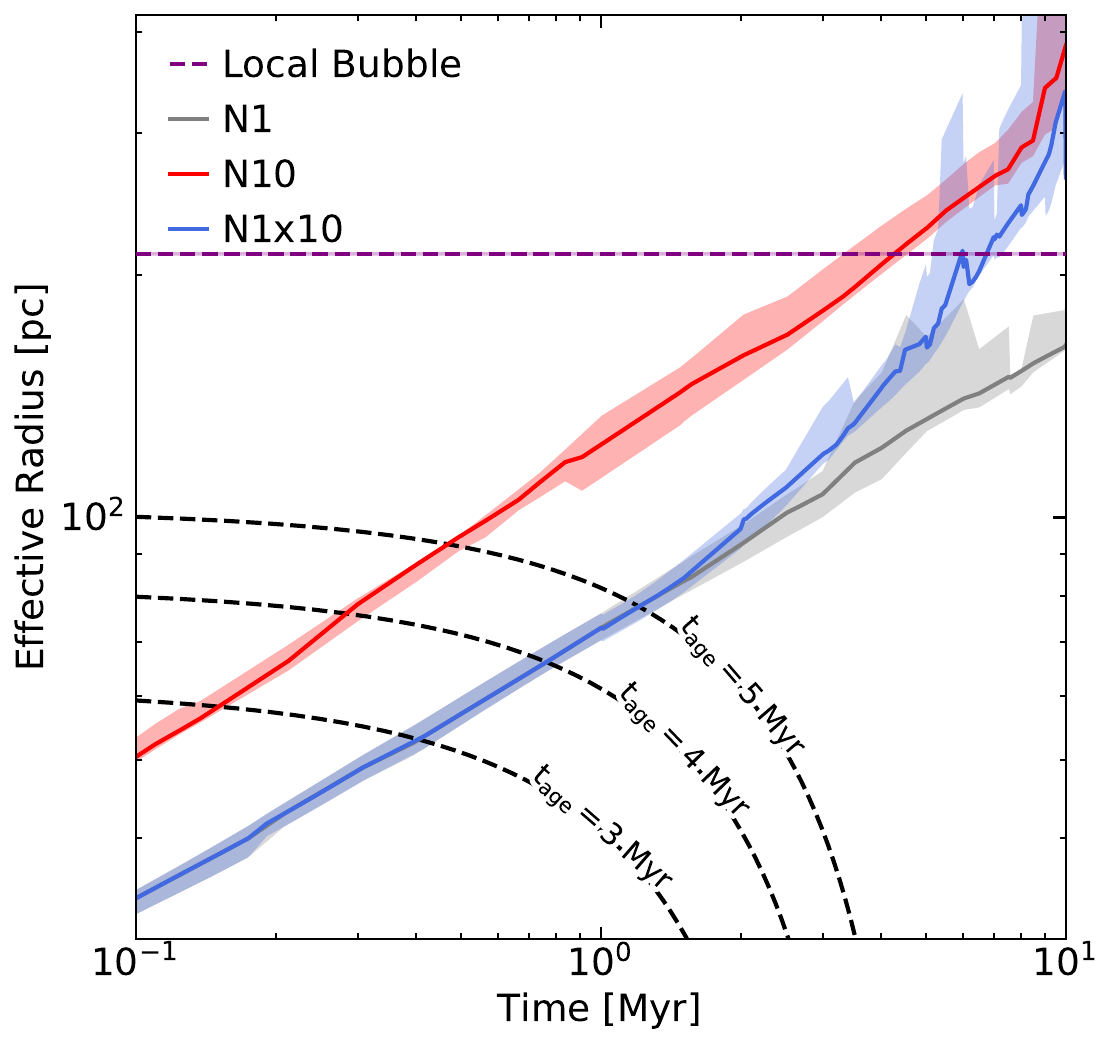}
 \caption{Crossing of the expanding LB's shell and the solar system.
 Gray, red, blue and purple lines are the same as in Fig. \ref{fig:size_evolution}.
 Black dashed lines correspond to the distance of the solar system from the center of the LB (Eq. \ref{eq:solar_system}) for different ages of the LB. 
 The solar system passes earlier through the LB's shell for more powerful explosions, but even for a single SN explosion it would have crossed the shell within the first $\sim 1\,\text{Myr}$ of its expansion.
  } 
 \label{fig:solar_system}
\end{figure}

Fossil records of sedimentary $^{60}$Fe and $^{244}$Pu on earth suggest an enrichment with SN ejecta over the past $4\,\text{Myr}$ \citep{2021Sci...372..742W}.
If this enrichment was due to the SNe powering the LB, the solar system must have entered the LB shortly before the enrichment started \citep{2016Natur.532...73B, 2023ApJ...947...58E, 2023A&A...680A..39S}.
Making use of this insight, \citetalias{2022Natur.601..334Z} estimated that the solar system has entered the LB $\sim 5\,\text{Myr}$ ago.

For a given expansion model for the LB's shell, we estimated when the solar system entered the LB, by comparing the trace-back distance of the solar system from the center of the LB to the radius of the expanding shell.
We adopted the same magnitude of $v_{\odot} \sim 20\,\text{km s}^{-1}$ for the speed of the solar system as
\citetalias{2022Natur.601..334Z} and assumed that it moves radially\footnote{We reconstructed the sun's orbit with \textsc{galpy} \citep{2015ApJS..216...29B} over the past 5 Myr and confirmed that this approximation is accurate to $\sim 1\, \%$.}, with a present-day distance from the center of $d_{\odot}\left(t_{\text{age}}\right) = 0$.

Under these assumptions, the distance of the solar system from the center of the LB is 
\begin{equation}\label{eq:solar_system}
    d_{\odot} \sim v_{\odot} \left(t_{\text{age}} - t\right) ~.
\end{equation}
In Fig. \ref{fig:solar_system} we show the passage of the solar system through the LB's shell.
For values of $t_{\text{age}} \gtrsim 4\,\text{Myr}$, the solar system would have crossed the LB's shell during the first $\lesssim 1\,\text{Myr}$ of its expansion.
Incorporation of sediments into earth's crust over the past $\sim 4\,\text{Myr}$ could have commenced shortly after the passage \citep{2023ApJ...947...58E, 2024ApJ...972..179E} and is therefore in agreement with the value $t_{\text{age}} \gtrsim 4\,\text{Myr}$ obtained above.

\section{Recent star formation and SNe in the solar neighborhood} \label{app:AlphaPersei}

\citetalias{2024Natur.631...49S} have used the star cluster catalogue of \citep{2023A&A...673A.114H} to group the clusters in the solar neighborhood into three distinct families of star clusters that likely share a common origin.
One of these cluster families, named $\alpha$Per, after its prominent member Alpha Persei can be both spatially and temporally associated with the LB.

Previous analyses of the origin of the LB usually focused on the role of Sco-Cen in driving its expansion \citep{2001ApJ...560L..83M, 2016Natur.532...73B, 2022Natur.601..334Z}. While the findings of \citetalias{2024Natur.631...49S} agree with these studies if only the SNe in Sco-Cen are considered, the potential contribution from other members of $\alpha$Per remains unclear.
\citetalias{2024Natur.631...49S} report the SFH in $\alpha$Per which suggests the existence of a peak in the star-formation activity associated with young clusters in Sco-Cen $\sim 10 \, \text{Myr}$ ago, which might be linked to a more recent burst in SN activity than previous studies considered.
Unfortunately, the low time resolution of the reported SFH does not allow to draw any further conclusions.

In order to investigate the role of this recent increase in star-formation, we used the data products of \citetalias{2024Natur.631...49S} to compute the recent SFH at higher temporal resolution and use it to estimate the SN-rate history for $\alpha$Per.
To this end, we used the 16\%, 50\% and 84\% quantiles of the cluster ages, masses and the number of SNe that exploded since their formation for all clusters associated with $\alpha$Per.

The SFH is the mass-weighted sum of the age distributions of each cluster $ \mathcal{P}_{i}\left(t\right)$, convolved with a window function $W\left(t; \Delta t\right)$:
\begin{equation}\label{eq:SFH}
    \text{SFH}\left(t\right) = \int_{-\infty}^{\infty} W\left(t - t_{\text{form}}; \Delta t\right) \sum_{i \in \text{family}} m_{i} \,  \mathcal{P}_{i}\left(t_{\text{form}}\right) \, \text{d}t_{\text{form}} ~. 
\end{equation}
Here $m_{i}$ is the mass of each cluster, which we sample from its posterior distribution.

We can reproduce the SFH reported by \citetalias{2024Natur.631...49S} if we use a top-hat filter with $\Delta t = 12.5\,\text{Myr}$ as the window function, use the median cluster mass for $m_{i}$ and use delta-peaks, peaked at the median age for the age distributions of the clusters $\mathcal{P}_{i}\left(t\right) = \delta\left(t - \text{median}\left(t_{\text{form, }i}\right)\right)$.
However, this treatment does not necessarily capture the sizeable uncertainties in the cluster ages and masses, encoded in their 16\% and 84\% quantiles.

To account for the uncertainty in the age and simultaneously push the time resolution of the SFH to the limit, we reconstructed plausible age distributions (Appendix~\ref{app:pdfs}) from the available data and set $W = \delta\left(t - t_{\text{form}}\right)$, namely, we took the SFH as the mass-weighted sum of the age distributions of the clusters.
We accounted for the uncertainty in the mass by sampling from the posterior distribution, leaving us with a number of sample SFHs. 

We estimated the SN-rate history, by assuming that the SN rate of each cluster remains at an approximately constant value between the onset of SN explosions $t_{\text{delay}}$ after the clusters formation and the life-time $t_{\text{active}}$ of the least massive stars undergoing type-II SN explosions \citep{1999ApJS..123....3L, 2017ApJ...834...25K} and is zero otherwise:
\begin{equation}\label{eq:SNR}
    \mathcal{R}_{\text{SN, }i}\left(t\right) = \frac{N_{\text{SN, }i}}{\text{min}\left(t_{\text{form, }i}, t_{\text{active}}\right) - t_{\text{delay}}} \, \mathbf{1}_{\left(t_{\text{delay}}, \, t_{\text{active}}\right)}\left(t_{\text{form, }i} - t\right) ~,
\end{equation}
where $\mathbf{1}_{X}\left(x\right)$ is the indicator function, which evaluates to unity if $x \in X$ and zero otherwise and $N_{\text{SN, }i}$ is the number of SNe that exploded since the formation of the cluster. We sampled both $N_{\text{SN, }i}$ and $t_{\text{form, }i}$ from their respective posterior distributions.

\subsection{Probability Distributions}\label{app:pdfs}

\begin{figure}
 \includegraphics[width=\linewidth, clip=true]{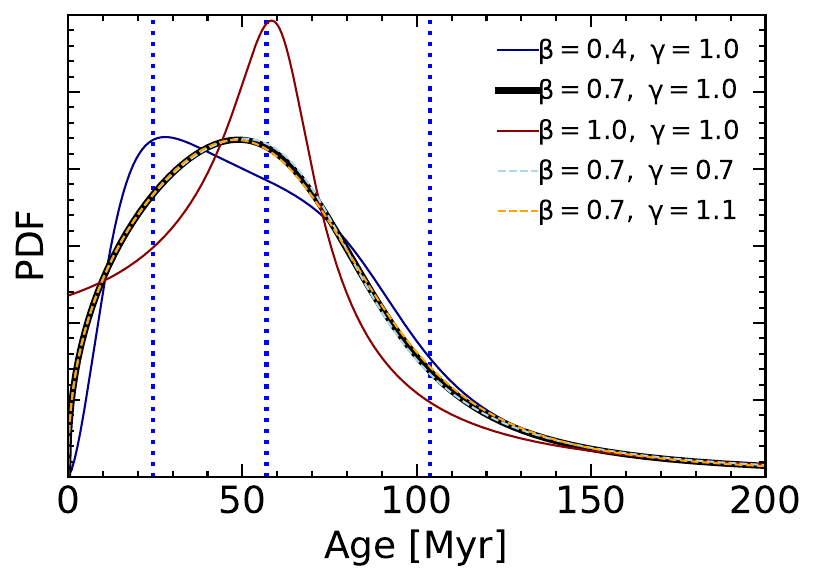}
 \caption{Reconstructed posterior age-pdf for the cluster ADS\_16795 for various values of $\beta$ and $\gamma$. Dotted, blue lines correspond to the 16\%, 50\% and 84\% quantiles.
 For smaller values of $\beta$ the mode of the distribution is closest to the 16\% quantile, while for $\beta \sim 1$ it is closer to the median. For $\beta \geq 1$ the pdf approaches a constant non-zero value for $x \rightarrow 0$.
 While the choice of $\gamma$ has little influence on the shape of the pdf, for $\gamma \gtrsim 1.1$ the pdf is not longer positive-definite. The pdf corresponding to our fiducial choice of parameters is depicted as a solid, black line.} 
 \label{fig:pdf}
\end{figure}

Given only a number of quantiles $x_{0} \leq ... \leq x_{i} \leq ... \leq x_{n}$ corresponding to percentiles $0=p_{0} < ... <p_{i}<...<p_{n}=1$ and a prior probability density function (pdf) $q\left(x\right)$, the most plausible posterior pdf is a locally reweighted version of the prior, where the weights are chosen such that the cumulative probabilities match those of the quantile constraints \citep{1957PhRv..106..620J}:
\begin{equation}\label{eq:max-ent}
    \mathcal{P}\left(x \left| q, \left(x_{i}\right)_{i=0,...,n}, \left(p_{i}\right)_{i=0,...,n} \right.\right) = q\left(x\right) \sum_{i=1}^n w_{i} \, \mathbf{1}_{\left(x_{i-1}, \,x_{i}\right)}\left(x\right) ~,
\end{equation}
with weights 
\begin{equation}
    w_{i} = \frac{p_{i} - p_{i-1}}{\mathbb{P}\left(x_{i-1} < x < x_{i}| q\right)} ~.
\end{equation}
With most smooth priors, the posterior distribution will exhibit large jumps, a feature that might be undesirable for our reconstruction of the SFH.

Generating smooth pdfs from quantile constraints is a common data-science application to which a number of creative approaches exist.
One particularly simple and flexible method that makes inverse-sampling trivial are so-called quantile-parametrized distributions \citep{2011DecA.8.206K}. 

The method uses a parameterization of the the quantile function that is linear in the quantile data constraints, using a number of arbitrary basis functions.
The corresponding pdf can then be simply derived using the inverse function rule of calculus, since the quantile function is the inverse of the cumulative distribution function.
However, care needs to be taken to ensure that the resulting pdf is positive definite.

We adopted this method to reconstruct the mass- and age posterior-pdfs using the following parameterization
\begin{equation}
    x\left(p\right) = \frac{p^\beta}{\left(1 - p\right)^\gamma} \left(a_{1} + a_{2}\,p + a_{3}\,\text{sin}\left(\pi\,p\right)\right) ~,
\end{equation}
where $\beta > 0$, $\gamma > 0$, $p \in [0, 1]$ is the cumulative probability, and $a_{i}$ are the parameters determined by the quantile constraints $x\left(p_{i}\right) = x_{i}$. For $\beta < 1$, the pdf tends towards zero for $y \rightarrow 0$, while for $\beta > 1$ it tends toward a constant value $> 0$.
We adopted $\gamma=1$ and $\beta = 0.7$ for which we confirmed that all pdfs are positive definite. 
In Fig. \ref{fig:pdf} we show how the pdf depends on the values of $\beta$ and $\gamma$ for one of the clusters in $\alpha$Per.

Since the number of SNe is limited to positive integers we did not need to worry about smoothness.
However, there are other complications. The data provided by \citetalias{2024Natur.631...49S} correspond to the quantiles of the marginal distribution. 
Yet, it is clear that the number of SNe has to correlated with the age of the cluster, at the very least, because it can only be non-zero for clusters older than $t_{\text{delay}}$.

To account for this correlation, we set the probability of $N_{\text{SN}}>0$ SNe to zero for $t_{\text{form}} < t_{\text{delay}}$, but otherwise kept the pdf for $N_{\text{SN}}>0$ unchanged with respect to the marginal distribution.
The marginal and joint probabilities of $N_{\text{SN}}=0$ SNe are then linked by the condition
\begin{equation}\label{eq:prob_null}
    \mathcal{P}\left(N_{\text{SN}} = 0\right) = p_{0} + \left(1-p_0\right)\,\mathcal{P}\left(N_{\text{SN}} = 0 \left|t_{\text{form}} > t_{\text{delay}}\right.\right) ~,
\end{equation}
where $p_{0} = \mathbb{P}\left(t_{\text{form}} < t_{\text{delay}}\right)$.

To define a sensible marginal pdf for the number of SNe, 
we assumed a constant prior for $N_{\text{SN}} < N_{\text{SN, }84\%}$, an exponential tail for $N_{\text{SN}} > N_{\text{SN, }84\%}$ and we set
\begin{equation}
    \mathcal{P}\left(N_{\text{SN}} = 0\right) = \max\left(p_{0}, \max_{\left\{i \, \left| \, Q_{i}= 0          
          \right.\right\}} p_i, \min_{\left\{i \, \left| \, Q_{i}> 0          
          \right.\right\}} \frac{p_{i} - p_{i-1}}{1 + Q_{i}}\right) ~,
\end{equation}
where $Q_{i}$ are the quantiles of the marginal distribution. 
The joint pdf of $t_{\text{form}}$ and $\text{N}_{\text{SN}}$ is then fully specified by applying Eqs. \ref{eq:max-ent} and \ref{eq:prob_null}. 

\subsection{SFH and SN-rate in $\alpha$Per}

\begin{figure}
 \includegraphics[width=\linewidth, clip=true]{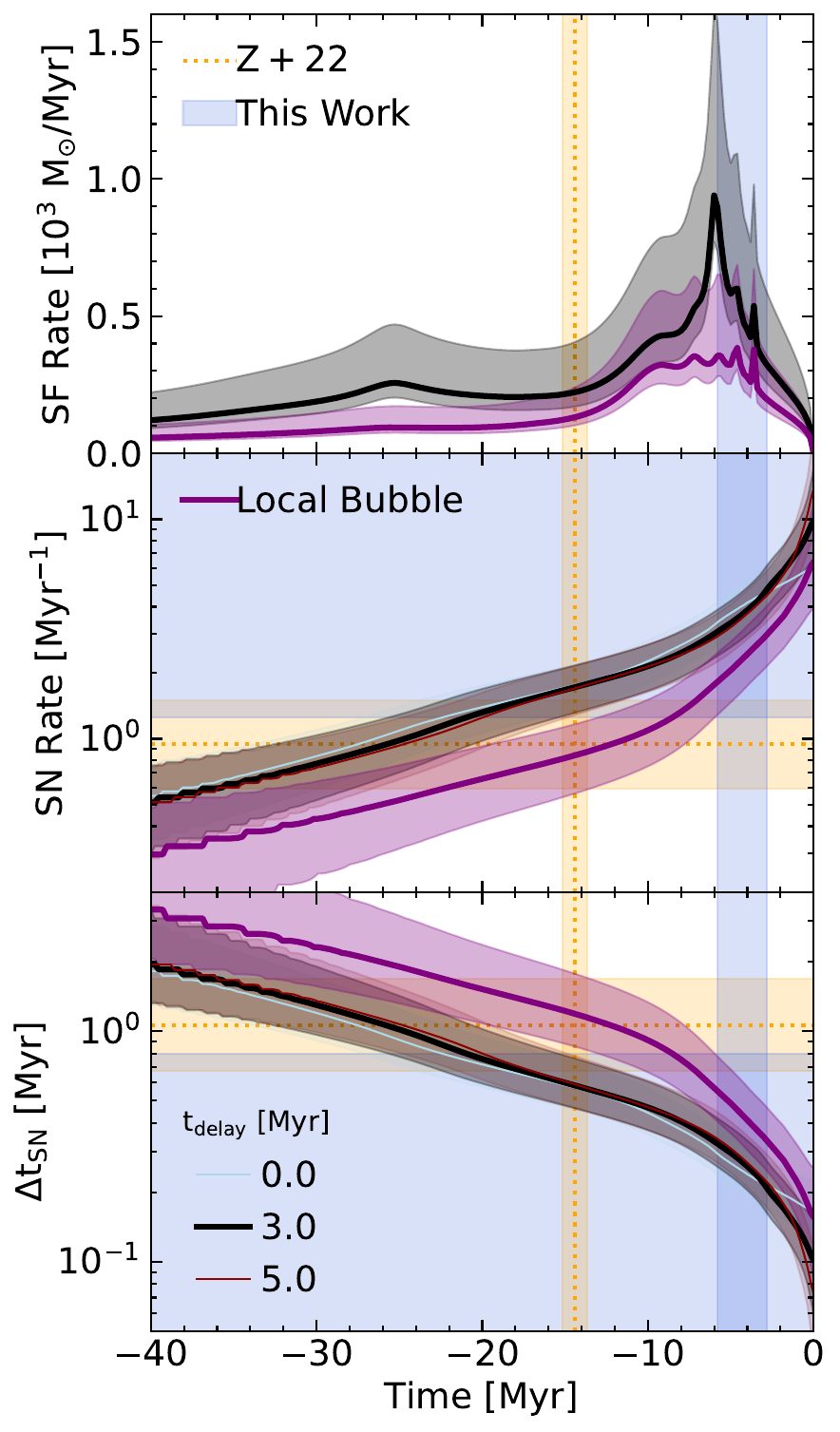}
 \caption{Top panel: Star-formation History of $\alpha$Per over the past 40 Myr. The SFR is peaked around 8 Myr ago. The peak is associated with recent star formation in Sco-Cen \citetalias{2024Natur.631...49S}. However, the dominant peak corresponds to the formation of the $\sim 1000\,\text{M}_{\odot}$ cluster Theia 38, at a distance of $\sim 500 \, \text{pc}$ from the solar system.
 Middle and bottom panel: SN rate and average time between SNe in $\alpha$Per over the past 40 Myr for different values of $t_{\text{delay}}$. The SN rate continuously grows (the time between SNe shortens) as the stellar mass of $\alpha$Per builds up, with a sharp increase in SN activity in the last few Myr, associated with the peak in the SFR, 8 Myr ago. The steepness of the increase slightly depends on the choice of $t_{\text{delay}}$, with a more gradual increase for $t_{\text{delay}} = 0$ and a steeper increase for $0 < t_{\text{delay}} < 8 \, \text{Myr}$.
The SFH, SN Rate and average time between SNe of the clusters spatially associated with the LB are shown in purple, where we assume a fiducial value of $t_{\text{delay}} = 3\,\text{Myr}$.
 Also shown as shaded regions are the age of the LB as well as the average time between SNe found by \citetalias{2022Natur.601..334Z} (orange) and this work (blue).} 
 \label{fig:SFH}
\end{figure}

\begin{figure*}
 \includegraphics[width=\linewidth, clip=true]{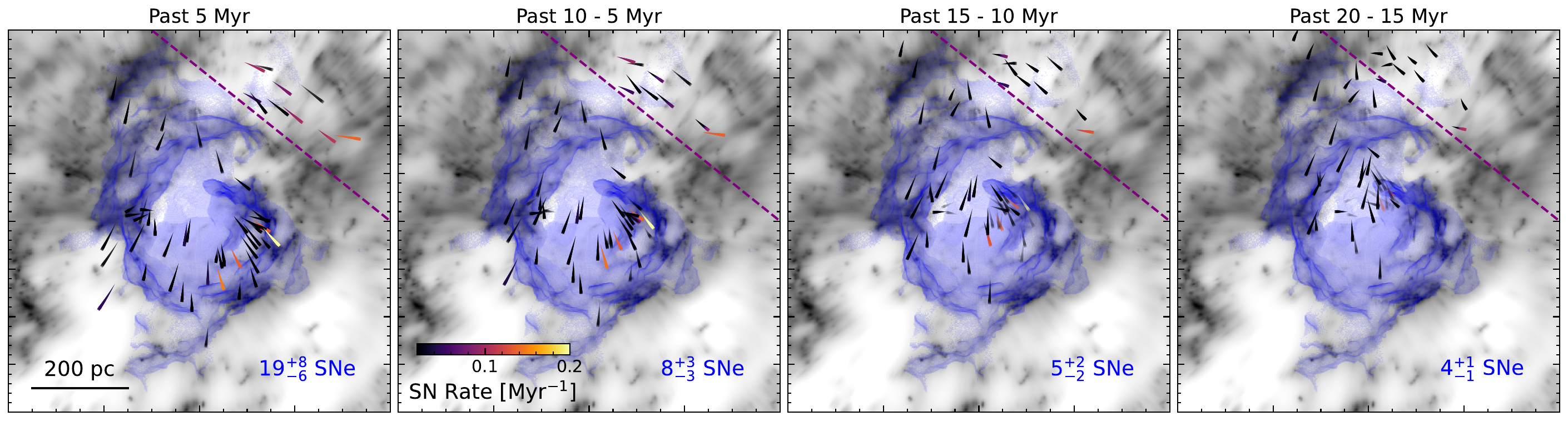}
 \caption{Trajectories of the member clusters of $\alpha$Per in the local standard of rest, within the vicinity of the LB, from left to right, over the past 5, $5-10$, $10-15$ and $15-20$ Myr. Trajectories are plotted as lines of increasing width, where increasingly wider lines correspond to later times. Trajectories are colored based on the clusters' instantaneous SN Rate and the opacity is set to the instantaneous probability that the cluster is between $t_{\text{delay}}$ and $t_{\text{active}}$ after its formation and is therefore potentially contributing SNe at a given time.
 While it is unlikely that a particular cluster colored black has contributed a SN, due to the large stochasticity in low mass clusters, the numerous black clusters might still collectively contribute a significant number of SNe. 
 The projected gas density as well as a projection of the LB's shell at the present time are also shown to provide context.
 We also show the number of SNe contributed within each time frame by the clusters, that are co-spatial with the current extent of the LB, using the purple dashed line to decide whether a cluster belongs to the LB or not. 
 With the exception of about 10 clusters in the top right (a few outside the frame) the majority of the clusters has been co-spatial with the current extent of the LB (in projection) for the past few Myr and therefore could have contributed SNe.
 The number of SNe contributed by nearby clusters in the past 5 Myr is consistent with our estimate in Tab. \ref{tab:SNe}.} 
 \label{fig:trajectories}
\end{figure*}

For each cluster in $\alpha$Per we drew a million samples of the cluster mass, age and number of SNe from their respective pdfs and used them to derive a million realizations of the SFH (Eq. \ref{eq:SFH}) and the SN-rates (Eq. \ref{eq:SNR}).
For each realization, we summed up the SN-rates of the clusters to obtain the SN-rate history of $\alpha$Per. We also derived the average time between SNe, by taking the reciprocal.

In Fig. \ref{fig:SFH} we show the 16\%, 50\% and 84\% quantile histories, obtained by drawing computing the quantiles at each point in time. We mark the onset of the expansion of the LB according to our model and that of \citetalias{2022Natur.601..334Z} as well as the SN rates, required by the respective models.

We find that there is a broad peak in the SFH $\sim 4-10 \, \text{Myr}$ ago, which according to \citetalias{2024Natur.631...49S} is associated with the formation of Upper Scorpius, Corona Australis, $\rho$ Ophiuchus, which are all part of Sco-Cen, and various other clusters in Taurus. 
While \citet{2023A&A...678A..71R} report a dominant peak in the SFH of Sco-Cen $\sim 15 \, \text{Myr}$ ago, the SFH in Fig. \ref{fig:SFH} associated with the LB is quite low $\lesssim 100 \, \text{M}_{\odot}\, \text{Myr}^{-1}$ before the onset of its peak $\sim 10\,\text{Myr}$.
These differences can likely be explained due to the different normalization (mass vs. star count), contributions from other clusters in our sample, and prominently the more accurate age estimates used by \citep{2023A&A...678A..71R}, further highlighting the need for accurate stellar ages.
Curiously, the range of ages for the LB found by our analysis above coincides with the downturn in the peak of star-formation activity, which is consistent with a scenario where the birth of the LB, powered by the delayed feedback from first stars formed in the peak, is quenching any further star-formation in the region, which is in stark contrast to the triggered star-formation scenario proposed by \citetalias{2022Natur.601..334Z}. This difference in interpretation results from the fact that the formation of the LB in the model of \citetalias{2022Natur.601..334Z} starts before the peak of the star formation activity while in our scenario it starts after.

The SFH can be compared with the SN rate, which has been steadily increasing from $\sim 1 \, \text{SN / Myr}$ 20 Myr ago to $\lesssim 10 \, \text{SNe / Myr}$ at the present day. The SN rate grows increasingly larger than the value of $\Delta t_{\text{SN}}^{-1} \sim 1 \,\text{Myr}^{-1}$ required by the model of \citetalias{2022Natur.601..334Z}, while it remains well within the range of values required by our analysis above.

To verify, whether this SN activity is actually spatially associated with the LB, in Fig. \ref{fig:trajectories} we show the trajectories of the clusters in $\alpha$Per in the vicinity of the LB, colored by their instantaneous SN Rate, over the past 5, $5-10$, $10-15$, and $15-20$ Myr.
Trajectories were calculated using the galactic dynamics package \textsc{galpy} \citep{2015ApJS..216...29B}, following the parameter choices outlined in \citetalias{2024Natur.631...49S}.
We find that the majority of the clusters has been co-spatial with the current extent of the LB over the past few Myr, with just over 10 of the 66 cluster members being too distant to have contributed.

In order to separate out the contributions from these distant clusters, we performed a simple spatial cut, indicated by the purple-dashed line in Fig. \ref{fig:trajectories}. 
The SN Rate of the remaining clusters is shown as a purple line in Fig. \ref{fig:SFH}, which shows that the SN Rate of the clusters co-spatial with the LB is still consistent with our findings, highlighted in Tab. \ref{tab:SNe}.

In Fig. \ref{fig:trajectories} we also show the number of SNe contributed by the clusters associated with the LB during each time frame.
Even though there were on average 9 SNe in the time window from 10-20 Myr ago, the fact that most of the contributing clusters have low SN rates suggests, that these SNe merged with the ISM in isolation before they could combine to form a coherent SB.
On the other hand, the more frequent SNe in the past 10 Myr might be more clustered, especially towards the Sco-Cen region, where 4-5 exceptionally active clusters reside.
These clustered SNe likely overlapped, forming the coherent LB that suppressed further star formation in the local ISM and reached a large enough size to be coherently powered by most subsequent SNe in the solar neighborhood.

\end{document}